\documentclass[aps, prd,preprintnumbers,superscriptaddress,nofootinbib,floatfix,notitlepage]{revtex4-1}

\usepackage[utf8]{inputenc}
\usepackage{graphicx} 
 \graphicspath{{./figures/}}
\usepackage{dcolumn} 
\usepackage{amsmath,amsfonts} 
\allowdisplaybreaks

\usepackage{mathabx}
\usepackage{physics}
\usepackage{slashed}
\usepackage{bm}
\usepackage{cancel}
\usepackage[caption=false]{subfig}
\usepackage{wrapfig}
\usepackage[dvipsnames,svgnames]{xcolor}
\usepackage{braket}
\usepackage{multirow}
\usepackage{graphicx}
\usepackage{hhline}
\usepackage[shortlabels]{enumitem}
\usepackage{float}
\usepackage{cancel}

\usepackage{hyperref} 

\usepackage[capitalise]{cleveref}

\hypersetup{%
  colorlinks=true, linktocpage=true, pdfstartpage=1, pdfstartview=FitV,%
  breaklinks=true, pdfpagemode=UseNone, pageanchor=true, pdfpagemode=UseOutlines,%
  plainpages=false, bookmarksnumbered, bookmarksopen=true, bookmarksopenlevel=1,%
  hypertexnames=true, pdfhighlight=/O,
  urlcolor=Cerulean, linkcolor=Cerulean, citecolor=Cerulean, 
}

\usepackage{framed}
\colorlet{shadecolor}{blue!20}
\usepackage[normalem]{ulem}

\begin{document}

\title{Testing Dark Matter Coscattering with the High-Energy LZ Recoil}
\author{Sreemanti Chakraborti}
\email{sreemanti.chakraborti@durham.ac.uk}
\affiliation{Institute for Particle Physics Phenomenology,
            Department of Physics,
            Durham University,
            Durham, DH1 3LE, United Kingdom}
\author{Soumyajit Datta}
\email{soumyajitdattaphys@gmail.com}
\affiliation{Indian Institute of Science Education and Research Kolkata, Mohanpur, 741246, India}

\begin{abstract}
    We investigate whether the high-energy nuclear recoil recently reported by LZ can probe dark matter (DM) whose relic abundance is set by coscattering, or conversion-driven freeze-out. We consider a singlet-doublet fermion DM model coupled to a light dark photon, in which the predominantly singlet-like DM state is separated from a more strongly interacting doublet-like state by an ${\cal O}(100)\,{\rm keV}$ mass splitting. In the small mixing regime, conversion into the doublet sector becomes the bottleneck controlling the relic abundance, realising coscattering, while the same small mixing would ordinarily suppress direct detection. A light dark photon enhances the endothermic transition $\chi_1 N\to\chi_2 N$, allowing the conversion-driven region to be probed by high-energy nuclear recoils. We find that for $m_{A'}=250~{\rm MeV}$ the LZ-sensitive region can cover essentially the full relic-compatible parameter space, including the dominantly conversion-driven regime. As the mediator mass is increased, the LZ sensitivity progressively shifts towards larger singlet-doublet mixing, reducing its overlap with the coscattering region. Combining the relic density and recoil requirements selects DM masses extending approximately from $750~{\rm GeV}$ to $1.6~{\rm TeV}$ for the mediator masses considered. The LZ recoil can therefore directly test the same interaction and physical parameters that dominantly set the DM relic abundance, providing a probe of the same conversion mechanism across vastly different cosmological epochs.
\end{abstract}
\preprint{IPPP/}

\maketitle

\section{Introduction}
\label{sec:intro}

Despite compelling gravitational evidence for dark matter (DM), its particle nature remains unknown. Weakly interacting massive particles (WIMPs) provide a particularly well-motivated framework, in which the observed relic abundance can arise from thermal freeze-out while the same interactions can be searched for in terrestrial experiments. Direct detection experiments have now reached unprecedented sensitivity to both conventional elastic scattering and scenarios with qualitatively different recoil signatures~\cite{Bertone:2004pz,Arcadi:2017kky,Arcadi:2024ukq,Liu:2017drf,Vahsen:2021gnb}.

Recently, the LUX-ZEPLIN (LZ) experiment has reported a single high-energy nuclear recoil event consistent with a recoil energy of $248\pm23\,({\rm stat})\pm23\,({\rm sys})~{\rm keV}$ in a $2.84$ tonne-yr exposure~\cite{LZ:2026axp}. The event is in tension with the background-only hypothesis at a global significance of $2.6\sigma$, reaching a maximal local significance of $3.4\sigma$. Although far from the $5\sigma$ discovery threshold, its occurrence in a low-background regime has motivated several DM interpretations, including Higgsino DM~\cite{Fan:2026kxx,Wu:2026nhi,Freese:2026sga,Du:2026guj, Bisal:2026khf,Frolovsky:2026tvq, Langhoff:2026ujr},
axion-portal DM~\cite{Visinelli:2026kgt,An:2026pkc,Unwin:2026rdp}, and other DM candidates~\cite{Yin:2026jnn,Smirnov:2026aqk,Nomura:2026qyq,Murayama:2026apt, Lou:2026idn, Su:2026rwz,Yamashita:2026ump,Chattopadhyay:2026ryw,deLima:2026shq,Wang:2026ytg,Dent:2026bji,McCabe:2026crm,Okada:2026upm,Ahmed:2026qjg,Gemmell:2026yaw,Okada:2026eol,Das:2026buc,De:2026win}.

The unusually large recoil energy makes inelastic DM particularly interesting. In such scenarios, the incident DM particle scatters into a slightly heavier state,
\begin{align}
    \chi^{\rm light} N \rightarrow \chi^{\rm heavy} N ,
\end{align}
with a small mass splitting between the two dark particles, $\chi^{\rm light}$ and $\chi^{\rm heavy}$ denote the light and heavy DM states, respectively, and $N$ denotes the target nucleus. The additional energy required for upscattering selects the high-velocity part of the Galactic DM distribution and can naturally shift the recoil spectrum towards larger energies. Studies of the connection between the LZ recoil and thermal inelastic DM with mass splittings of ${\cal O}(100)\,{\rm keV}$ include Ref.~\cite{DiMauro:2026ldr,Su:2026rwz}. Complementary interpretations have considered non-DM explanations, such as
neutrino-induced events~\cite{Brdar:2026ukx,Jeesun:2026vzo,
Chattaraj:2026fxn} or disappearance of a bound
neutron~\cite{Aghaie:2026vsu}.\footnote{See~\cite{Cyncynates:lzmap} for a steadily expanding list of LZ event interpretations. We hope to join the list soon!} The Higgsino interpretation has also been examined using complementary constraints from solar capture and neutrino searches~\cite{Pospelov:2026ewn,Bose:2026ndd,Nguyen:2026lui,DiMauro:2026dqp,Rodd:2026tyn}.

In the context of inelastic DM as an explanation of the excess, a simple framework in which such a compressed spectrum arises is the singlet-doublet dark matter (SDDM) model~\cite{Bhattacharya:2017sml,Paul:2024prs,Bhattiprolu:2026hnp}. The dark sector contains an electroweak singlet fermion and a vector-like electroweak doublet which mix after symmetry breaking. The resulting lightest neutral state can be predominantly singlet-like, while a nearby neutral state retains predominantly doublet-like interactions.

Several recent works have explored singlet-doublet realisations of the reported LZ event~\cite{Bandyopadhyay:2026gjw,Borah:2026zwf,Lee:2026jxl}, demonstrating that a singlet admixture provides a natural way of retaining an inelastic transition while suppressing otherwise excessive electroweak scattering rates. Related constructions have further emphasised the role of singlet-doublet mixing in suppressing the inelastic transition rate and relaxing solar-capture constraints, while simultaneously obtaining a thermal relic abundance. 

While the complementarity between direct detection and relic density has
already been demonstrated by the above references, the connection explored
in this work is different. We focus on the small-mixing regime in which conversion between the predominantly singlet and doublet states becomes the bottleneck
controlling the relic abundance, rather than merely participating in the
thermal freeze-out dynamics. This conversion-driven or coscattering
regime therefore makes the relic abundance intrinsically sensitive to the
interaction structure responsible for the off-diagonal
$\chi^{\rm light}\leftrightarrow\chi^{\rm heavy}$ transition. In coscattering, the DM abundance freezes out when bath-assisted inelastic
conversion between the light and heavy dark states becomes inefficient~\cite{DAgnolo:2017dbv,Garny:2017rxs,Brummer:2019inq,Liu:2025swd,Cheng:2018vaj,DiazSaez:2024nrq,Belanger:2025wjh,Junius:2019dci}.
The mechanism is therefore intrinsically inelastic, making it particularly
well suited to connect the thermal history to an inelastic direct detection
signal. The small mixing required to enter
this regime would ordinarily make the corresponding direct detection signal
difficult to observe. A light dark photon changes this situation, allowing
the suppressed off-diagonal transition to generate an observable
endothermic $\chi^{\rm light} N\to\chi^{\rm heavy} N$ rate while preserving the conversion-driven thermal history. Dark photon-mediated endothermic scattering has been previously explored in connection with the LZ recoil~\cite{Zhu:2026dag}. In that case, however, the relic abundance is determined by secluded annihilation into the dark sector, independently of the inelastic transition responsible for direct detection. In the singlet-doublet framework considered here, the mixing provides an additional control parameter that allows this transition itself to become the bottleneck governing the relic abundance.

While the ${\cal O}(100)\,{\rm keV}$
splitting is practically irrelevant for freeze-out, as also pointed out in previous studies of thermal inelastic DM~\cite{DiMauro:2026ldr,Bandyopadhyay:2026gjw}, and primarily determines
the kinematics of the LZ upscattering, the dark gauge coupling,
singlet-doublet mixing and mediator mass govern both the conversion
dynamics in the early Universe and the strength of the present-day
inelastic transition.

We determine the parameter space in which the observed DM abundance is
obtained through this conversion-driven dynamics and investigate whether it
can simultaneously yield an ${\cal O}(1)$ event rate in the high energy LZ
recoil region. We find a restricted region in which the LZ-sensitive
parameter space overlaps directly with that in which conversion controls
the relic abundance. The relic abundance and LZ event thus probe the same
underlying physical parameters and off-diagonal interaction structure at
widely separated cosmological epochs, precisely because the early Universe
abundance is controlled by coscattering.

This paper is organised as follows. In Sec. \ref{sec:model}, we introduce the minimal framework responsible for the generation of inelastic DM. In Sec.~\ref{sec:dm}, we discuss the DM relic abundance and identify the
parameter space in which coscattering is realised. In
Sec.~\ref{sec:dd}, we study the resulting inelastic nuclear recoil signal
and its connection to the high-energy LZ event. We discuss additional
experimental constraints in Sec.~\ref{sec:other_constraints}, before discussing the results of the numerical analysis in Sec.~\ref{sec:bound} and finally 
concluding in Sec.~\ref{sec:con}.

%
\section{Minimal framework}
\label{sec:model}
%
We are interested in dark sectors containing two nearly degenerate states, $\chi_1$ and $\chi_2$, for which interactions with the Standard Model (SM) can induce the inelastic transition:
\begin{equation}
    \chi_1 + {\rm SM} \longleftrightarrow \chi_2 + {\rm SM}.
\end{equation}
A simple realisation is obtained when the two states couple off-diagonally to a light vector mediator $A'$, which in turn communicates with the SM through kinetic mixing. At low energies the relevant interactions can be written as:
\begin{align}
    \mathcal{L} \supset
    g_{12} A'_\mu
    \left(
        \bar{\chi}_2 \gamma^\mu \chi_1
        + \bar{\chi}_1 \gamma^\mu \chi_2
    \right)
    + \epsilon e A'_\mu J_{\rm EM}^{\mu},
    \label{eq:}
\end{align}
where $g_{12}$ denotes the transition coupling and $\epsilon$ parametrises the coupling of $A'$ to the electromagnetic current.

To realise this structure, we consider a vector-like Dirac fermionic singlet $\xi=(\xi_L,\xi_R)$, a vector-like Dirac fermionic doublet
$\Psi=(\Psi_L,\Psi_R)$ with $\Psi_{L,R}=(\psi^0,\psi^-)_{L,R}^{T}$, and a complex scalar singlet $\Phi$ charged under a dark $U(1)_D$. Their quantum numbers under $SU(2)_L\times U(1)_Y\times U(1)_D$ are:

\begin{align}
    \xi_{L,R} &\sim (\mathbf{1},0)_0,
    \qquad
    \Psi_{L,R} \sim (\mathbf{2},-1/2)_1,
    \qquad
    \Phi \sim (\mathbf{1},0)_1.
\end{align}

where the subscript denotes the $U(1)_D$ charge. We further impose a $\mathbb{Z}_2$ symmetry under which $\xi$ and $\Psi$ are odd, while all SM fields and the dark sector bosons are even. The lightest $\mathbb{Z}_2$-odd state is therefore stable and
constitutes the DM candidate.

 The singlet and doublet fermions are connected through the effective operators:
\begin{align}
    \mathcal{L} \supset
    -\left[
        \frac{c_R}{\Lambda}\Phi\,\bar{\Psi}_L \tilde{H} \xi_R
        + \frac{c_L}{\Lambda}\Phi\,\bar{\Psi}_R \tilde{H} \xi_L
        + {\rm h.c.}
    \right],
\end{align}

where $H$ is the SM Higgs doublet and $\widetilde{H}=i\sigma_2H^*$. $c_L$ and $c_R$ are dimensionless Wilson coefficients and $\Lambda$ denotes the scale of the effective interaction. A renormalisable completion generating the operators above is discussed in Appendix~\ref{app:uv}. 

The dark gauge sector contains:
\begin{align}
    \mathcal{L} \supset
    -\frac{1}{4}X_{\mu\nu}X^{\mu\nu}
    -\frac{\epsilon}{2}B_{\mu\nu}X^{\mu\nu}
    + |D_\mu\Phi|^2
    - V(H,\Phi).
\end{align}

where $X_{\mu\nu}$ denotes the $U(1)_D$ field-strength tensor and $B_{\mu\nu}$ the SM hypercharge field-strength tensor. For $\Phi$, the covariant derivative is:
\begin{align}
    D_\mu = \partial_\mu - ig_DX_\mu.
\end{align}

The scalar potential $V(H,\Phi)$ is taken to be:

\begin{align}
    V(H,\Phi)
    =
    -\mu_H^2 H^\dagger H
    +\lambda_H(H^\dagger H)^2
    -\mu_D^2\Phi^\dagger\Phi
    +\lambda_D(\Phi^\dagger\Phi)^2
    +\lambda_{HD}(H^\dagger H)(\Phi^\dagger\Phi).
\end{align}
In our analysis we set the Higgs-portal coupling $\lambda_{HD}=0$, such that the SM and dark Higgs bosons do not mix.

After electroweak and $U(1)_D$ symmetry breaking one obtains:
\begin{align}
    H =
    \frac{1}{\sqrt{2}}
    \begin{pmatrix}
        0 \\ v+h
    \end{pmatrix},
    \qquad
    \Phi = \frac{v_D+h_D}{\sqrt{2}},
\end{align}
where $v$ and $v_D$ denote the electroweak and dark sector VEVs, respectively, while $h$ and $h_D$ are the corresponding
physical scalars. Taking $c_L=c_R\equiv c$, the neutral fermion mass matrix is:
\begin{align}
    \mathcal{M}_0 =
    \begin{pmatrix}
        M_\xi & m \\
        m   & M_\Psi
    \end{pmatrix},
    \qquad
    m = \frac{c\,v\,v_D}{2\Lambda}.
\end{align}
We define the dark fermionic mass eigenstates by
\begin{align}
    \begin{pmatrix}
        \xi \\ \psi^0
    \end{pmatrix}
    =
    \begin{pmatrix}
        c_\theta & s_\theta \\
        -s_\theta &  c_\theta
    \end{pmatrix}
    \begin{pmatrix}
        \chi_1 \\ \chi_2
    \end{pmatrix},
    \qquad
    \tan 2\theta = \frac{2m}{M_\Psi-M_\xi},
\end{align}
where $s_\theta\equiv\sin\theta$ and $c_\theta\equiv\cos\theta$. 

We take $m_{\chi_1}<m_{\chi_2}$ and work predominantly in the small-mixing regime, where $\chi_1$ is singlet-like and $\chi_2$ doublet-like. We define
\begin{align}
    \delta \equiv m_{\chi_2}-m_{\chi_1},
    \qquad
    m = \delta\,s_\theta c_\theta.
\end{align}
Since only the doublet $\Psi$ among the dark fermions carries $U(1)_D$ charge, its gauge interaction generates both diagonal and off-diagonal couplings in the mass basis:
\begin{align}
    \mathcal{L} \supset -g_D A'_\mu \big[
        &s_\theta^2\,\bar{\chi}_1\gamma^\mu\chi_1
        +c_\theta^2\,\bar{\chi}_2\gamma^\mu\chi_2
        \nonumber\\
        &-s_\theta c_\theta
        \left(
            \bar{\chi}_1\gamma^\mu\chi_2
            + {\rm h.c.}
        \right)
    \big],
\end{align}
and hence
\begin{align}
    g_{12}=g_D s_\theta c_\theta.
\end{align}
The structure assumed above therefore follows directly from singlet-doublet mixing. Notice in particular that for $s_\theta\ll1$, the diagonal $\chi_1$ coupling scales as $s_\theta^2$, while the inelastic transition coupling
scales as $s_\theta$.

The breaking of $U(1)_D$ also gives:
\begin{align}
    m_{A'}=g_D v_D,
    \qquad
    m_{h_D}^2=2\lambda_D v_D^2.
\end{align}
%
\vspace{2em}
Expanding the same operator responsible for the neutral-fermion mixing gives:
\begin{align}
    \mathcal{L} \supset
    -\frac{m}{v_D}h_D \big[
        &-2s_\theta c_\theta
        \left(
            \bar{\chi}_1\chi_1-\bar{\chi}_2\chi_2
        \right)
        \nonumber\\
        &+(c_\theta^2-s_\theta^2)
        \left(
            \bar{\chi}_1\chi_2+{\rm h.c.}
        \right)
    \big],
\end{align}
while the dark Higgs kinetic term contains:
\begin{align}
    \mathcal{L} \supset
    \frac{m_{A'}^2}{v_D}h_D A'_\mu A'^\mu
    +\frac{m_{A'}^2}{2v_D^2}h_D^2 A'_\mu A'^\mu.
\end{align}
Together, the interactions above specify the dark sector interactions relevant below.

We parameterise the mass of the charged dark fermionic component, $\psi^\pm$, relative to the predominantly doublet-like neutral state $\chi_2$ as:
\begin{align}
    m_{\psi^\pm}=m_{\chi_2}+\Delta_C,
\end{align}
where $\Delta_C$ denotes the dark charged-neutral mass splitting. Electroweak radiative corrections generate a positive charged-neutral mass splitting of $\sim 350~{\rm MeV}$ for a predominantly doublet-like state. Corrections associated with singlet-doublet mixing are negligible in the small-mixing regime considered here. We therefore take $\Delta_C\simeq 350~\mathrm{MeV}$ as a representative value throughout the analysis.

The low-energy parameter space can therefore be expressed as:
\begin{align}
    \left\{
        m_{\chi_1},\,\delta,\,s_\theta,\,
        g_D,\,m_{A'},\,m_{h_D},\,\epsilon,\,\Delta_C
    \right\}.
\end{align}
\section{Dark Matter Coscattering and Inelastic Conversion}
\label{sec:dm}
%
The off-diagonal interactions introduced in the previous section play a
dual role. For direct detection, they facilitate the inelastic transition:
\begin{align}
    \chi_1 + {\rm SM} \to \chi_2 + {\rm SM},
\end{align}
while in the early Universe the same interactions mediate conversions
between the predominantly singlet-like DM state and the heavier,
predominantly doublet-like state:
\begin{align}
    \chi_1 + {\rm SM/Dark} \leftrightarrow
    \chi_2 + {\rm SM/Dark}.
\end{align}
where ``Dark" denotes scattering with particles in the dark sector bath. Thus, although the kinematics and cosmological environment are very
different, both processes probe the same underlying off-diagonal
interaction structure.

For small singlet-doublet mixing, i.e., when $\sin \theta \ll 1 $, annihilation of $\chi_1$ is strongly suppressed. Its abundance can instead be controlled by conversions into the heavier state $\chi_2$, whose doublet component allows the odd sector to remain efficiently coupled to the thermal bath. If the rate of these conversion processes becomes insufficient compared with the Hubble expansion while the heavier states are still efficiently depleted, the total dark sector abundance departs from equilibrium and the relic density becomes controlled by the freeze-out of the conversion process rather than by annihilation of $\chi_1$ itself. This realises the coscattering, or the conversion-driven freeze-out regime.

Importantly, coscattering does not necessarily imply an immediate loss of chemical equilibrium between $\chi_1$ and $\chi_2$. Additional dark sector interactions can continue to efficiently redistribute
the abundances of the two odd states even as the combined odd sector begins to depart from equilibrium with the thermal bath. In this regime one may therefore have:
\begin{align}
    \frac{Y_1}{Y_{1,\rm eq}}
    \simeq
    \frac{Y_2}{Y_{2,\rm eq}}
    \neq 1,
\end{align}
while the $1\,0\leftrightarrow2\,0$ conversion processes remain crucial for transferring the odd-sector population into the efficiently depleted doublet-like sector and thereby controlling the relic abundance.

To describe this regime, it is useful to separate the particle content according to its thermal evolution. We follow the treatment introduced in Ref.~\cite{Paul:2024prs} and assign the particles in thermal equilibrium with the bath to sector $0$, while the two populations of $\mathbb{Z}_2$-odd states are tracked as independent sectors,
\begin{align}
    0 &= \{ {\rm SM},\, A',\, h_D \},
    \nonumber\\
    1 &= \{ \chi_1 \},
    \nonumber\\
    2 &= \{ \chi_2,\,\psi^\pm \}.
    \label{eq:thermal_sectors}
\end{align}
The separation of $\chi_1$ and the predominantly doublet-like states into different sectors is essential in the conversion-driven regime, since chemical equilibrium between them cannot be assumed a priori, even though it may remain approximately maintained in parts of the parameter space considered below.

The reactions relevant for the thermal evolution can then be classified according to the sectors of their initial and final states. We denote a generic process by $a b \to c d$, where $a,b,c,d\in\{0,1,2\}$ label the corresponding thermal sectors.  Several qualitatively different classes of reactions enter the coupled evolution.

The first class consists of direct annihilation of the DM,
\begin{align}
    1\,1 \leftrightarrow 0\,0,
    \label{eq:1100}
\end{align}
corresponding to processes of the form
$\chi_1\chi_1\to {\rm bath}$.  Since $\chi_1$ is predominantly singlet-like, these rates are suppressed by powers of the small singlet-doublet mixing parameter $\sin\theta$.

A second class contains annihilation and co-annihilation processes involving the heavier, predominantly doublet-like states,
\begin{align}
    1\,2 &\leftrightarrow 0\,0,
    \label{eq:1200}\\
    2\,2 &\leftrightarrow 0\,0.
    \label{eq:2200}
\end{align}
The latter include annihilations of $\chi_2$ and $\psi^\pm$, as well as their mutual co-annihilations.  Contrary to $\chi_1$, these states inherit the electroweak interactions of the vector-like doublet and can therefore be depleted efficiently even for very small $\sin\theta$. Dark sector annihilation channels involving $A'$ and $h_D$ provide additional depletion modes as the dark gauge coupling $g_D$ is increased. The $2\,2\leftrightarrow 0\,0$ reactions consequently provide the efficient sink required for conversion-driven freeze-out.

Of particular importance are processes which transfer particles between sectors $1$ and $2$ without changing the total number of $\mathbb{Z}_2$-odd particles,
\begin{align}
    1\,0 \leftrightarrow 2\,0.
    \label{eq:1020}
\end{align}
These are the inelastic conversion, or coscattering reactions. They can proceed through either the SM or dark components of the thermal bath. For an SM bath particle $f$, the $A'$-mediated process takes the form:
\begin{align}
    \chi_1 f \leftrightarrow \chi_2 f ,
    \label{eq:SM_conversion}
\end{align}
with the product of couplings entering the matrix element scaling as:
\begin{align}
    g_{A'\chi_1\chi_2}\,g_{A' f f}
    \sim
    \epsilon e Q_f\,g_D s_\theta c_\theta .
    \label{eq:SM_conversion_coupling}
\end{align}
Electroweak interactions inherited from the doublet component generate additional SM-assisted conversion processes.  These reactions realise in the early Universe the same underlying off-diagonal
$\chi_1\leftrightarrow\chi_2$ transition that is probed by inelastic direct
detection, as we discuss in the following sections.

The presence of $A'$ and $h_D$ in the thermal bath opens an additional set of purely dark conversion reactions,
\begin{align}
    \chi_1 A' &\leftrightarrow \chi_2 h_D ,
    \nonumber\\
    \chi_1 h_D &\leftrightarrow \chi_2 A' .
    \label{eq:dark_conversion}
\end{align}
These processes belong to the same $1\,0\leftrightarrow2\,0$ class as Eq.~\eqref{eq:SM_conversion}; the distinction is whether the bath particle participating in the conversion is an SM or a dark sector state. Their rates involve the following off-diagonal interactions
\begin{align}
    |g_{A'\chi_1\chi_2}|
        &= g_D s_\theta c_\theta ,
    \nonumber\\
    |g_{h_D\chi_1\chi_2}|
        &= \frac{m}{v_D}
           |c_\theta^2-s_\theta^2|
         = \frac{\delta}{v_D}s_\theta c_\theta
           |c_\theta^2-s_\theta^2| ,
    \label{eq:dark_conversion_couplings}
\end{align}
together with the corresponding diagonal interactions.  Since $v_D=m_{A'}/g_D$, the relative importance of the dark bath conversion channels increases with the dark gauge coupling $g_D$. Consequently,
for fixed $\epsilon$ and mediator masses, the dominant contribution to the
$1\,0\leftrightarrow2\,0$ conversion rate can shift from SM-assisted
processes at smaller $g_D$ to processes involving $A'$ and $h_D$ at larger
$g_D$. This changes the composition of the total conversion rate, while the
underlying mechanism remains $1\,0\leftrightarrow2\,0$ coscattering.

In addition to bath-assisted conversions, reactions involving only the
$\mathbb{Z}_2$-odd sectors can redistribute the abundances between sectors
$1$ and $2$. The relevant reaction classes include
\begin{align}
    1\,1 &\leftrightarrow 2\,2,
    &
    1\,2 &\leftrightarrow 2\,2,
    &
    1\,2 &\leftrightarrow 1\,1 .
    \label{eq:dark_redistribution}
\end{align}
These processes therefore enter the
coupled Boltzmann evolution of $Y_1$ and $Y_2$ given below.

Defining $Y_i=n_i/s$, where $n_i$ denotes the total number density of particles belonging to sector $i$ and $s$ is the entropy density, we follow the sector decomposition introduced above and write

\begin{align}
\frac{dY_1}{dT}
={}&
\frac{1}{3H}\frac{ds}{dT}
\Bigg[
\langle\sigma_{1100}v\rangle
\left(Y_1^2-Y_{1}^{\rm eq\,2}\right)
\nonumber\\
&+
\langle\sigma_{1122}v\rangle
\left(
Y_1^2-
Y_2^2\frac{Y_{1}^{\rm eq\,2}}{Y_{2}^{\rm eq\,2}}
\right)
\nonumber\\
&+
\langle\sigma_{1200}v\rangle
\left(
Y_1Y_2-Y_{1}^{\rm eq}Y_{2}^{\rm eq}
\right)
\nonumber\\
&+
\langle\sigma_{1222}v\rangle
\left(
Y_1Y_2-
Y_2^2\frac{Y_{1}^{\rm eq}}{Y_{2}^{\rm eq}}
\right)
\nonumber\\
&-
\langle\sigma_{1211}v\rangle
\left(
Y_1Y_2-
Y_1^2\frac{Y_{2}^{\rm eq}}{Y_{1}^{\rm eq}}
\right)
\nonumber\\
&-
\frac{\Gamma_{2\rightarrow1}}{s}
\left(
Y_2-
Y_1\frac{Y_{2}^{\rm eq}}{Y_{1}^{\rm eq}}
\right)
\Bigg],
\label{eq:BE_Y1}
\\[2mm]
\frac{dY_2}{dT}
={}&
\frac{1}{3H}\frac{ds}{dT}
\Bigg[
\langle\sigma_{2200}v\rangle
\left(Y_2^2-Y_{2}^{\rm eq\,2}\right)
\nonumber\\
&-
\langle\sigma_{1122}v\rangle
\left(
Y_1^2-
Y_2^2\frac{Y_{1}^{\rm eq\,2}}{Y_{2}^{\rm eq\,2}}
\right)
\nonumber\\
&+
\langle\sigma_{1200}v\rangle
\left(
Y_1Y_2-Y_{1}^{\rm eq}Y_{2}^{\rm eq}
\right)
\nonumber\\
&-
\langle\sigma_{1222}v\rangle
\left(
Y_1Y_2-
Y_2^2\frac{Y_{1}^{\rm eq}}{Y_{2}^{\rm eq}}
\right)
\nonumber\\
&+
\langle\sigma_{1211}v\rangle
\left(
Y_1Y_2-
Y_1^2\frac{Y_{2}^{\rm eq}}{Y_{1}^{\rm eq}}
\right)
\nonumber\\
&+
\frac{\Gamma_{2\rightarrow1}}{s}
\left(
Y_2-
Y_1\frac{Y_{2}^{\rm eq}}{Y_{1}^{\rm eq}}
\right)
\Bigg].
\label{eq:BE_Y2}
\end{align}
Here $\langle\sigma_{\alpha\beta\gamma\delta}v\rangle$ denotes the
thermally averaged cross section for reactions
$\alpha\,\beta\leftrightarrow\gamma\,\delta$, summed over the particles
belonging to the corresponding thermal sectors.  The equilibrium
abundances are denoted by $Y_i^{\rm eq}$.

The effective rate $\Gamma_{2\rightarrow1}$ collects processes which convert a sector-$2$ particle into a sector-$1$ particle. It contains both decays,
\begin{align}
    2\rightarrow1+0,
\end{align}
and the bath-assisted $2\,0\leftrightarrow1\,0$ coscattering processes
discussed above. Schematically,
\begin{align}
    \Gamma_{2\rightarrow1}
    =
    \Gamma_{2\rightarrow1}^{\rm dec}
    +
    \Gamma_{2\rightarrow1}^{\rm scat}.
    \label{eq:Gamma21}
\end{align}
The decay contribution accounts for unstable sector-$2$ states
eventually converting into the stable sector-$1$ state and bath
particles.  In particular, for the compressed neutral spectrum
$\chi_2$ ultimately decays into $\chi_1$ through the electroweak
interactions inherited from its doublet component.

Lastly, when conversion between the two odd sectors is sufficiently rapid, their abundances may remain in chemical equilibrium. If, in addition, the combined odd sector remains efficiently coupled to the thermal bath through annihilation and co-annihilation processes, the system follows the usual co-annihilation evolution~\cite{Griest:1990kh}, with the total abundance freezing out once these number-depleting reactions become inefficient.

We solve Eqs.~\eqref{eq:BE_Y1} and \eqref{eq:BE_Y2} using
\texttt{micrOMEGAs}~\cite{Belanger:2026asz}, keeping sectors $1$ and $2$ independent rather
than imposing chemical equilibrium.  The calculation therefore
includes annihilation, co-annihilation, dark sector redistribution,
decay and bath-assisted conversion processes within a common coupled
Boltzmann evolution, and continuously interpolates between the
co-annihilation and conversion-driven regimes.

\section{Dark Matter Phenomenology}
\label{sec:pheno}
%
\subsection{High-energy nuclear recoils at LZ}
\label{sec:dd}
The off-diagonal dark photon interaction introduced in Sec~\ref{sec:model} mediates the endothermic scattering of the lighter state off the detector nuclei $N$ as:
\begin{align}
    \chi_1 + N \rightarrow \chi_2 + N ,
\end{align}
where
\begin{align}
    \delta \equiv m_{\chi_2}-m_{\chi_1}>0 .
\end{align}
In our model, the inelastic transition can be mediated by the $Z$, Higgs and the dark photon $A'$. For the parameter region of interest, however, the nuclear recoil rate is dominated by $A'$ exchange, in particular for $m_{A'}\sim\mathcal{O}(100~{\rm MeV})$. Neglecting the subleading electroweak contributions, the differential cross section for scattering off a Xe isotope with nuclear mass $m_N$ is
\begin{align}
    \frac{d\sigma_N}{dE_R}
    =
    \frac{
        2m_N Z^2
        \left(\epsilon e g_D s_\theta c_\theta\right)^2
    }{
        \pi v^2
        \left(m_{A'}^2+2m_NE_R\right)^2
    }
    F_A^2(q),
    \qquad
    q^2=2m_NE_R .
    \label{eq:inelastic_xsec}
\end{align}
where $Z=54$ for Xe and $F_A(q)$ denotes the nuclear form factor. $v$ denotes the DM velocity in the galactic halo and $q$ represents the momentum transferred in the scattering.

Unlike elastic scattering, the incoming DM particle must provide the additional energy required to produce the heavier state $\chi_2$. The minimum incoming velocity required for a recoil $E_R$ is therefore~\cite{XENON:2022avm}

\begin{align}
    v_{\min}(E_R)
    =
    \frac{1}{\sqrt{2m_NE_R}}
    \left(
        \frac{m_NE_R}{\mu_{\chi N}}+\delta
    \right),
    \qquad
    \mu_{\chi N}
    =
    \frac{m_{\chi_1}m_N}{m_{\chi_1}+m_N}.
    \label{eq:vmin_inelastic}
\end{align}

The minimum of $v_{\min}$ occurs at

\begin{align}
    E_R^\star
    =
    \frac{\mu_{\chi N}}{m_N}\,\delta,
    \qquad
    v_{\rm thr}
    =
    \sqrt{\frac{2\delta}{\mu_{\chi N}}}.
    \label{eq:inelastic_threshold}
\end{align}

where $E_R^\star$ is the recoil energy at which $v_{\min}(E_R)$
is minimised, while $v_{\rm thr}$ is the corresponding minimum
incoming DM speed required for the endothermic scattering to occur.

Consequently, increasing $\delta$ simultaneously raises the
velocity threshold and shifts the kinematically preferred recoil energy towards
larger $E_R$. This feature is particularly relevant for the isolated
high-energy nuclear-recoil candidate reported by LZ at
$E_R\simeq248~{\rm keV}$.

We calculate the recoil spectrum by summing over the naturally occurring Xe isotopes as given below:
\begin{align}
    {}^{124}{\rm Xe},\,
    {}^{126}{\rm Xe},\,
    {}^{128}{\rm Xe},\,
    {}^{129}{\rm Xe},\,
    {}^{130}{\rm Xe},\,
    {}^{131}{\rm Xe},\,
    {}^{132}{\rm Xe},\,
    {}^{134}{\rm Xe},\,
    {}^{136}{\rm Xe},
\end{align}

weighted by their natural isotopic abundances. In particular, the largest abundances are carried by ${}^{129}{\rm Xe}$, ${}^{131}{\rm Xe}$ and ${}^{132}{\rm Xe}$, with fractions $0.264$, $0.212$ and $0.269$, respectively~\cite{Raj:2024guv}. The isotope dependence is retained both in the reduced mass entering Eq.~\eqref{eq:vmin_inelastic} and in the nuclear form factor. For the latter we use the standard Helm parametrisation~\cite{ENGEL1991114,Vietze:2014vsa},

\begin{align}
    F_A(q)
    =
    3\,\frac{j_1(qR_1)}{qR_1}
    \exp\left[-\frac{(qs)^2}{2}\right],
    \label{eq:helm}
\end{align}

with

\begin{align}
    R_1^2
    &=
    c_A^2+\frac{7\pi^2}{3}a^2-5s^2,
    &
    c_A&=1.23A^{1/3}-0.60~{\rm fm},
    &
    a&=0.52~{\rm fm},
    &
    s&=0.9~{\rm fm}.
\end{align}

The nuclear suppression is substantial at the recoil energy of the LZ event. For example, at $E_R=248~{\rm keV}$ we find

\begin{align}
    F_{129}^2 &\simeq 2.6\times10^{-4},
    &
    F_{131}^2 &\simeq 1.7\times10^{-4},
    &
    F_{132}^2 &\simeq 1.3\times10^{-4}.
    \label{eq:helm_248}
\end{align}

The corresponding momentum transfer is $q\simeq245~{\rm MeV}$. Thus, although the high-energy recoil lies within the kinematically accessible region, coherent Xe scattering is already strongly suppressed by loss of nuclear coherence.

The differential recoil rate is obtained from~\cite{Cooley:2021rws}

\begin{align}
    \frac{dR}{dE_R}
    =
    \frac{\rho_\chi}{m_{\chi_1}}
    \sum_A N_A
    \int_{v>v_{\min}^A}
    d^3v\,
    f_{\rm lab}(\mathbf v)\,
    v\,
    \frac{d\sigma_A}{dE_R},
    \label{eq:recoil_rate}
\end{align}

where $f_{\rm lab}(\mathbf v)$ is the dark matter velocity distribution in the lab frame, $\rho_{\chi}$ is the local DM energy density, and $N_A$ denotes the number of target nuclei of isotope $A$ per unit detector mass, including its natural isotopic abundance. To describe $f_{\rm lab}(\mathbf v)$, we adopt the Standard Halo Model (SHM)~\cite{Baxter:2021pqo,LEWIN199687} with $v_0=220~{\rm km\,s^{-1}}$, $v_{\rm esc}=544~{\rm km\,s^{-1}}$, and
$v_E=232~{\rm km\,s^{-1}}$.

\begin{figure}[htb!]
    \centering
    \includegraphics[width=0.8\textwidth]{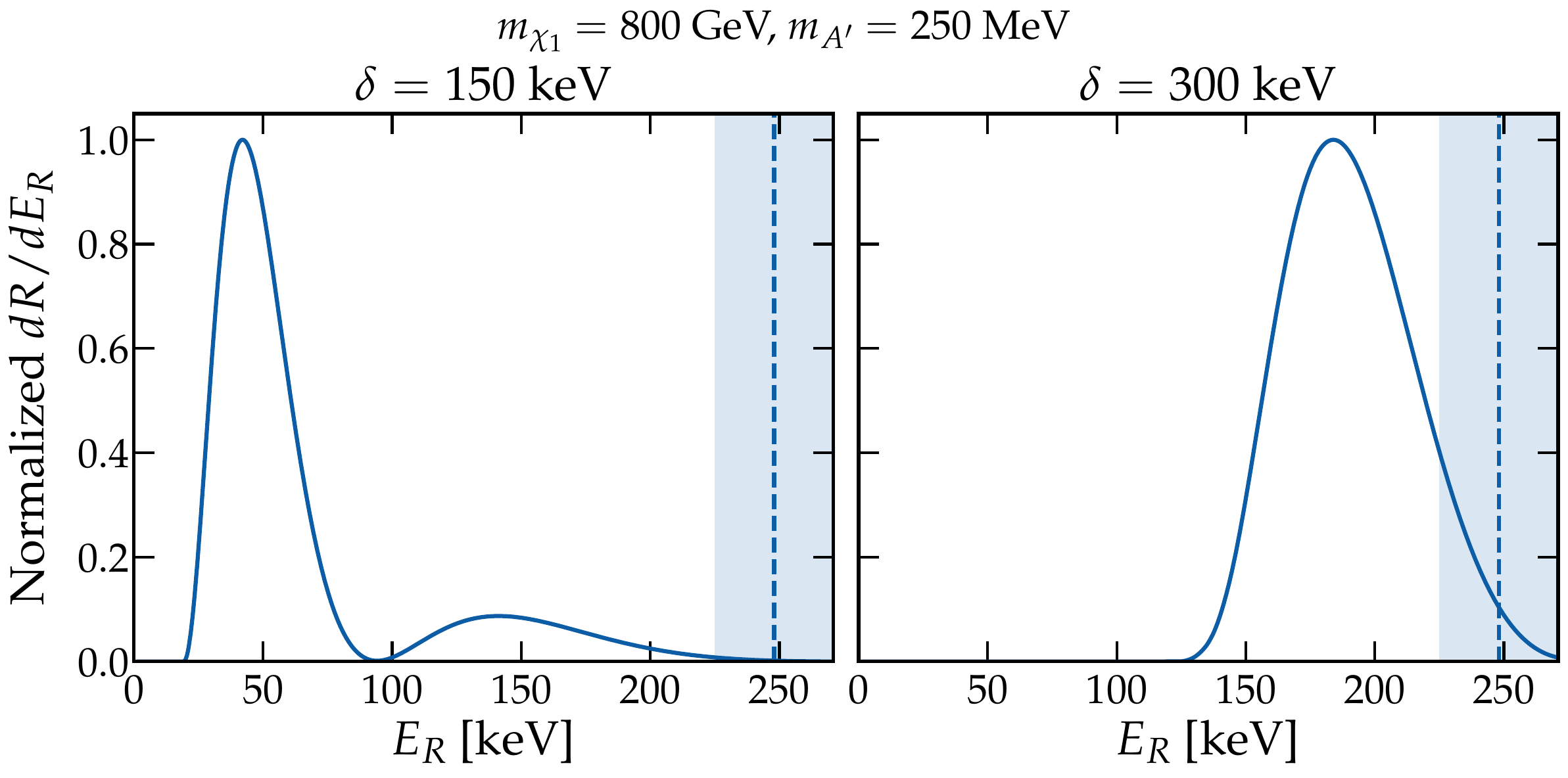}
    \caption{Normalised Xe nuclear-recoil spectrum for
    $m_{\chi_1}=800~{\rm GeV}$ and $m_{A'}=250~{\rm MeV}$.
    The {\it left} and {\it right} panels correspond to
    $\delta=150~{\rm keV}$ and $300~{\rm keV}$, respectively.
    The dashed vertical line indicates the $E_R\simeq248~{\rm keV}$ high-energy LZ event, while the shaded
    region denotes the $\pm 23~{\rm keV}$ statistical recoil energy interval around the event.}
    \label{fig:recoil}
\end{figure}

The role of the inelastic splitting is illustrated in
Fig.~\ref{fig:recoil} where the normalised recoil spectrum\footnote{The two-lobed structure visible for the smaller splitting originates from the first diffraction minimum of the Xe Helm form factor around $E_R\sim 90$--$100~{\rm keV}$. For larger $\delta$, the increasing inelastic velocity threshold suppresses the low-energy lobe, leaving
the higher energy recoil region dominant.} is plotted as a function of $E_R$ for two representative values of $\delta$. For the choice of
$m_{\chi_1}=800~{\rm GeV}$, $m_{A'}=250~{\rm MeV}$ and a relatively
small splitting, $\delta=150~{\rm keV}$, we find that the spectrum is still dominated by substantially lower recoil energies, with the dominant peak around $\sim 40~{\rm keV}$. The event at
$248~{\rm keV}$ then lies far out on the suppressed high-energy tail,
where both the halo integral and the Helm form factor reduce the rate.
Increasing the splitting shifts the kinematically preferred recoil
region upwards. For $\delta=300~{\rm keV}$ the spectrum is instead
concentrated at ${\cal O}(150$--$200)~{\rm keV}$, leaving appreciable recoil rate in the vicinity of the high-energy LZ event.

There is nevertheless an upper kinematic limit on the splitting. For a given recoil energy and maximum available lab-frame DM speed, 
the largest accessible splitting follows from Eq.~\ref{eq:vmin_inelastic} as
\begin{align}
    \delta_{\max}(E_R;v_{\max})
    =
    v_{\max}\sqrt{2m_NE_R}
    -
    \frac{m_NE_R}{\mu_{\chi N}}.
    \label{eq:delta_max}
\end{align}
Taking $v_{\max}$ as the sum of the galactic escape velocity ($v_{\rm esc}$) and Earth's velocity in the galactic frame ($v_E$), i.e.,  $v_{\max}=v_{\rm esc}+v_E\simeq776~{\rm km\,s^{-1}}$, the
representative mass number $A=131$, $m_{\chi_1}=800~{\rm GeV}$ case gives
\begin{align}
    \delta_{\max}(E_R=248~{\rm keV})
    \simeq 351~{\rm keV}.
\end{align}
The high-energy recoil therefore favours a characteristic window of
mass splittings of order a few hundred keV: splittings that are too
small leave the event deep in the high-energy tail of the recoil
spectrum, whereas splittings approaching $\delta\simeq350~{\rm keV}$ encounter the endpoint of the available halo phase space. This motivates the approximate range taken in our analysis:
\begin{align}
    \boxed{
        \delta \sim 150\text{--}300~{\rm keV}
    }
    \label{eq:delta_LZ_range}
\end{align}

\subsection{Accelerator constraints}
\label{sec:other_constraints}

The light dark sector states are subject to complementary constraints from accelerator searches. In the parameter region considered here, the most relevant bounds arise from searches for visible decays of dark photon ~\cite{BaBar:2014zli,LHCb:2017trq} and from dark Higgstrahlung~\cite{McKenna:2018xjb}. For the mass hierarchy $m_{h_D}\simeq2.4\,m_{A'}$ adopted here, the dark Higgs predominantly decays into a pair of dark photons, $h_D\to A'A'$, making visible dark Higgstrahlung at $e^+e^-$ colliders directly relevant.

Kinetic mixing induces the coupling
\begin{align}
    {\cal L}\supset \epsilon e A'_\mu J^\mu_{\rm EM},
\end{align}
allowing $A'$ to be produced through $e^+e^-\to\gamma A'$ and subsequently decay into kinematically accessible charged SM states. Searches for narrow resonances in these visible final states constrain the $(m_{A'},\epsilon)$ plane. Throughout our analysis we fix $\epsilon=10^{-4}$. For the mediator masses considered here, $m_{A'}=250$, $500~{\rm MeV}$ and $1~{\rm GeV}$, this choice remains compatible with the corresponding visible dark photon searches
~\cite{BaBar:2014zli,LHCb:2017trq}.

A complementary constraint is provided by dark Higgstrahlung,
\begin{align}
    e^+e^-\rightarrow A'^{*}\rightarrow A'h_D,
    \qquad
    h_D\rightarrow A'A',
\end{align}
leading to a three dark photon final state. Unlike single-$A'$ production,
the corresponding production rate depends also on the dark gauge coupling,
and scales schematically as
\begin{align}
    \sigma(e^+e^-\to A'h_D)
    \propto \epsilon^2\alpha_D,
    \qquad
    \alpha_D=\frac{g_D^2}{4\pi}.
\end{align}
For the value of kinetic mixing adopted here this combination becomes
\begin{align}
    \epsilon^2\alpha_D
    \simeq 8.0\times10^{-10}\,g_D^2 ,
\end{align}
and hence remains below $8.0\times10^{-10}$ over the range
$g_D\leq1$ considered in our scan. Visible dark Higgstrahlung searches at the $B$ factories probe precisely this combination and therefore provide a
particularly relevant constraint towards the large-$g_D$ boundary of our parameter space~\cite{McKenna:2018xjb,Jaegle:2015fme,BaBar:2012bkw}. For our benchmark mass ratios and $\epsilon=10^{-4}$, the parameter region considered below remains compatible with the corresponding limits.

These accelerator constraints are complementary to the cosmological and direct detection requirements considered in the analysis. In particular, the small kinetic mixing suppresses production through the visible sector portal, while $g_D$ can remain sufficiently large for dark sector
conversion to control the relic abundance. This separation allows the light dark photon to enhance the present-day inelastic scattering rate without excluding the conversion-driven region through existing accelerator searches.

\section{Results}
\label{sec:bound}

\begin{figure}[htb!]
    \centering
    \includegraphics[width=0.85\textwidth]{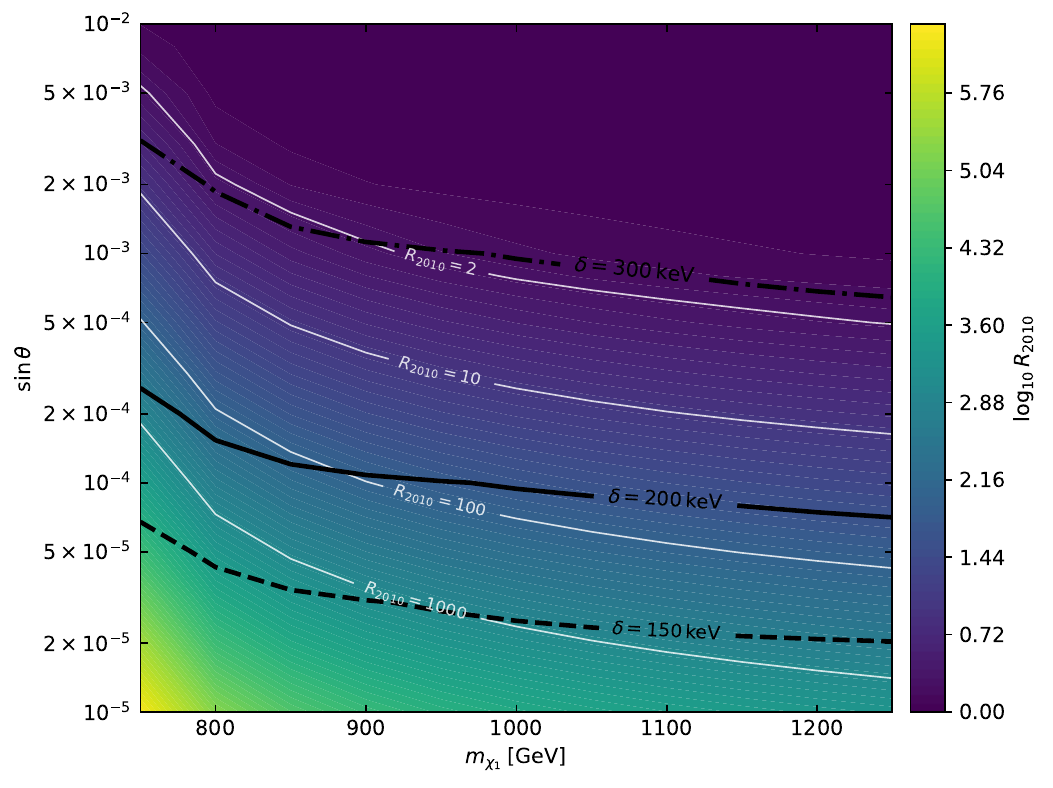}
    \caption{Relic density and LZ sensitivity in the $(m_{\chi_1},\sin\theta)$
plane for $m_{A'}=250~{\rm MeV}$ and $\epsilon=10^{-4}$. At each point,
$g_D$ is chosen to reproduce the observed DM relic abundance. The colour
scale shows $\log_{10}R_{2010}$, where $R_{2010}$ quantifies the change in
the relic abundance when the conversion processes $10\leftrightarrow20$
are removed; larger values therefore indicate an increasingly
conversion-driven regime. The white contours show representative values
of $R_{2010}$. The black curves correspond to one expected LZ event,
$N_{\rm LZ}=1$, for mass splittings $\delta=150$, $200$, and
$300~{\rm keV}$, as indicated.}
    \label{fig:combined}
\end{figure}

In this section we combine the relic density requirement with the high-energy LZ recoil. In Fig.~\ref{fig:combined}, we show the resulting parameter space in the $(m_{\chi_1},\sin\theta)$ plane for three representative dark photon masses below, denoting $m_{A'}$ fixed at 250~MeV, 500~MeV and 1~GeV. We also fix $\epsilon=10^{-4}$ and choose $m_{h_D}/m_{A'}\simeq 2.4$, such that $h_D\to A' A'$ remains kinematically open.

For each point, $g_D$ is adjusted to reproduce the observed total DM relic abundance. The mixing angle is varied over $10^{-5}\leq\sin\theta\leq10^{-2}$ in order to capture the region in which the relic density and direct detection requirements can simultaneously be relevant. The dashed, solid and dot-dashed black bands indicate the regions yielding $N_{\rm LZ}\simeq1$ for three representative inelastic splittings, $\delta=150$, $200$ and $300~{\rm keV}$, respectively,\footnote{For these small splittings, we have checked that the loop-induced dipole decay $\chi_2\to\chi_1\gamma$, mediated by the charged
$\psi^\pm$--$W^\mp$ loop, occurs sufficiently late to convert the post-freeze-out $\chi_2$ abundance into $\chi_1$ without affecting the cosmological evolution relevant here.} in the high-energy recoil window, while the white solid lines denote contours of the knockout ratio of the conversion process,
\begin{align}
    \mathcal{R}_{2010}
    =\frac{\Omega_{\neg 2010}h^2}{\Omega_{\rm full}h^2}
    \simeq\frac{\Omega_{\neg 2010}h^2}{0.12},
\end{align}
where $\Omega_{\neg 2010}h^2$ is the relic density obtained without the
$2010$ processes. The colour scale shows $\mathcal{R}_{2010}$, with larger
values indicating an increasingly important role of the
$2\,0\leftrightarrow1\,0$ conversion process in setting the relic density.

An important feature of the combined parameter space is the emergence of a lower bound on $m_{\chi_1}$. At smaller masses, the electroweak interactions of the doublet-like sector provide very efficient conversion between the singlet- and doublet-like sectors. Schematically, these processes take the form
\begin{align}
    D + {\rm SM} \leftrightarrow \chi_1 + {\rm SM}
\end{align}
where $D$ denotes either of the doublet-like states ($\psi^\pm$ or $\chi_2$). Their rates are controlled by the electroweak interactions together with the singlet-doublet mixing. 

As a consequence, the relic abundance becomes too small even in the limit of very weak dark sector interactions, $g_D\lesssim10^{-2}$, where dark bath conversion is
effectively switched off. The correct relic abundance can then only be recovered by decreasing $\sin\theta$, thereby suppressing the communication between the singlet-like state and the electroweak sector. However, the same reduction, combined with the necessary small $g_D$, suppresses the inelastic dark photon coupling relevant for LZ,
\begin{align}
    g_{A'\chi_1\chi_2}\simeq g_D\sin\theta ,
\end{align}
and consequently removes the high-energy recoil signal. We therefore do not display this low-mass region, where the relic density and LZ requirements cannot be simultaneously realised.

However, these contributions become less efficient as $m_{\chi_1}$ increases. This is because at fixed $\sin\theta$, and once $m_{\chi_1}$ sets the dominant mass scale, the corresponding thermally averaged cross sections scale parametrically as
\begin{align}
    \langle\sigma v\rangle_{\rm EW\,conv}
    \sim \frac{g_{\rm EW}^4\sin^2\theta}{m_{\chi_1}^2}
\end{align}
up to threshold, kinematic and thermal effects. Since freeze-out occurs at approximately fixed $x=m_{\chi_1}/T$, the interaction rate for scattering with relativistic bath particles scales as
\begin{align}
    \Gamma_{\rm EW\,conv}
    \sim T^3\langle\sigma v\rangle_{\rm EW\,conv},
\end{align}
whereas $H\sim T^2/M_{\rm Pl}$. Consequently,
\begin{align}
    \frac{\Gamma_{\rm EW\,conv}}{H}
    \sim
    \frac{g_{\rm EW}^4\sin^2\theta\,M_{\rm Pl}}{m_{\chi_1}},
\end{align}
again up to the detailed thermal functions and thresholds. The electroweak conversion network therefore becomes progressively less efficient relative to the Hubble expansion as the overall DM mass scale is increased. This explains why its effect is strongest toward the low-mass edge of the viable region.

The situation changes around  $m_{\chi_1}\sim 750~{\rm GeV}$. For sufficiently large $g_D$, the dark bath conversion processes $\chi_1 A'\leftrightarrow\chi_2 h_D$ and $\chi_1 h_D\leftrightarrow\chi_2 A'$ become the dominant conversion channels controlling the transfer between the two odd sectors. The relic abundance then enters the conversion-driven regime discussed in Sec.~\ref{sec:dm}, while the larger dark coupling also makes the off-diagonal interaction relevant for the LZ recoil sufficiently
strong.

Along the relic density contour, the value of $g_D$ required to maintain the conversion-driven depletion increases with $m_{\chi_1}$. At larger masses, the conversion-depletion chain becomes less efficient at fixed couplings; therefore, a progressively stronger
dark sector interaction is required to maintain sufficient conversion of $\chi_1$ into the efficiently depleted heavier sector. The corresponding hierarchy and evolution of the conversion and depletion rates are discussed in Appendix~\ref{app:kinetic_equilibrium}. This behaviour continues until $g_D$ reaches the upper value adopted in our scan, $g_D=1$, which determines the high-$m_{\chi_1}$ endpoint of the displayed parameter space.

As for the LZ bands, a larger $\delta$ raises the minimum velocity required for the high-energy recoil as explained in Sec.~\ref{sec:dd}, thereby reducing the available halo phase space. Maintaining $N_{\rm LZ}\sim 1$ therefore requires a larger off-diagonal coupling $g_D\sin\theta$, which within the relic-compatible parameter space shifts the corresponding region toward larger $\sin\theta$.

\begin{figure}[htb]
    \centering
    \includegraphics[width=0.49\textwidth]{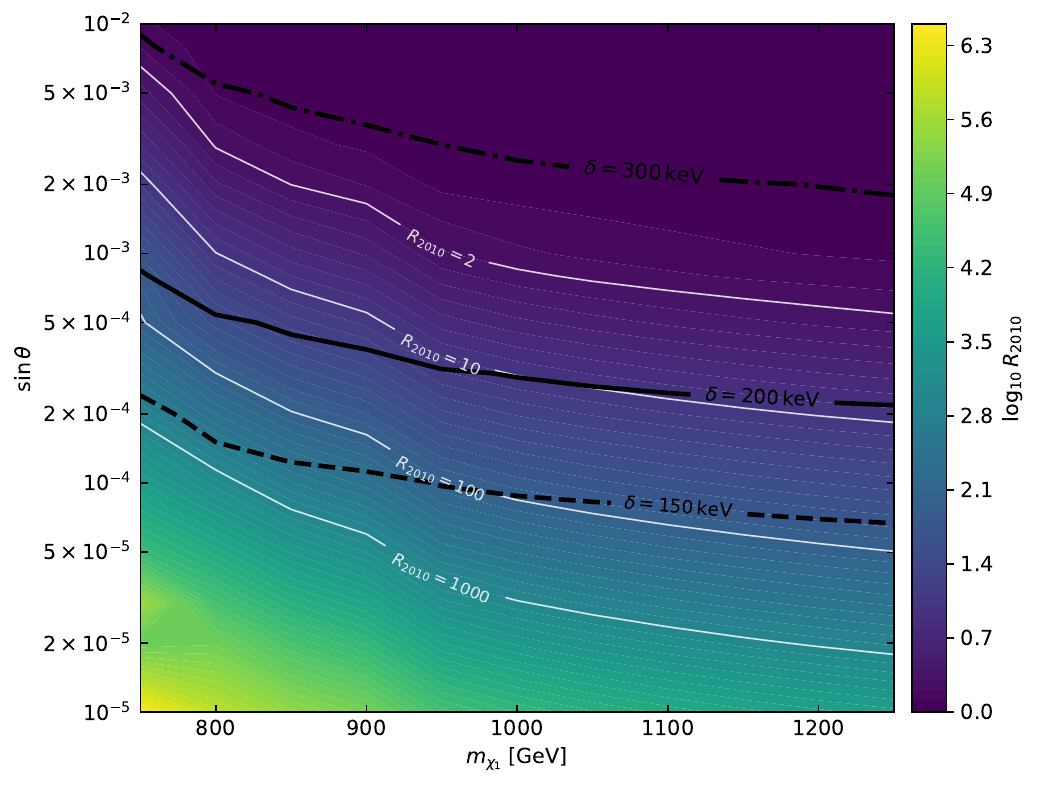}
    \hfill
    \includegraphics[width=0.49\textwidth]{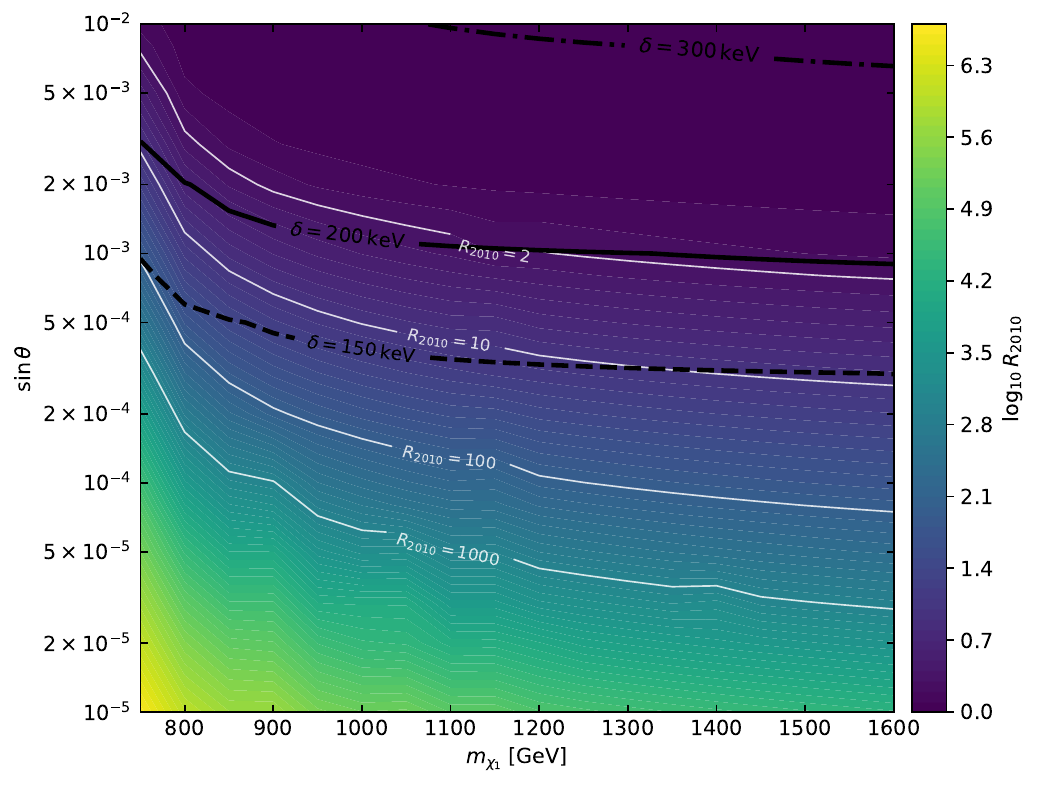}
    \caption{Same as Fig.~\ref{fig:combined}, but for $m_{A'}=500~{\rm MeV}$
({\it left}) and $m_{A'}=1~{\rm GeV}$ ({\it right}). Increasing the mediator mass
suppresses the inelastic scattering rate, shifting the $N_{\rm LZ}=1$
contours towards larger singlet-doublet mixing. Consequently, the overlap
between the LZ-sensitive and conversion-driven regions is progressively
reduced, particularly for the larger mass splittings. For
$m_{A'}=1~{\rm GeV}$, the relic-compatible parameter space extends to
larger DM masses, while the $\delta=300~{\rm keV}$ LZ contour no longer
overlaps with the strongly conversion-driven region.}
    \label{fig:combined2}
\end{figure}

The two figures in Fig.~\ref{fig:combined2} show the combined reach for larger mediator masses. Increasing the dark photon mass to $m_{A'}=500~{\rm MeV}$ leaves the relic density behaviour qualitatively unchanged. The region in which the $2\,0\leftrightarrow1\,0$ conversion becomes important, as well as the corresponding range of $m_{\chi_1}$, remains similar to the $250~{\rm MeV}$ case. The LZ bands, however, are shifted towards larger mixing angles. This follows directly from the suppression of the inelastic scattering amplitude by the heavier mediator,
\begin{align}
    \frac{d\sigma_A}{dE_R}\propto
    \frac{g_D^2\sin^2\theta\cos^2\theta}
    {(m_{A'}^2+q^2)^2}.
\end{align}
Since the relevant momentum transfer is $q\sim245~{\rm MeV}$, increasing $m_{A'}$ from $250$ to $500~{\rm MeV}$ already leads to a substantial propagator suppression. A larger $\sin\theta$ is therefore required to obtain the same $N_{\rm LZ}\simeq1$ rate.

For $m_{A'}=1~{\rm GeV}$, the relic density behaviour changes more noticeably. The correct relic abundance can be obtained with smaller values of $g_D$, such that the $g_D=1$ boundary is reached only at larger $m_{\chi_1}$ and the viable parameter space consequently extends to higher DM masses. At the same time, the stronger mediator suppression of the direct detection rate pushes the LZ bands further towards larger mixing angles. This increasingly separates the LZ-sensitive region from the conversion-driven regime for larger inelastic splittings. In particular, the $\delta=300~{\rm keV}$ band no longer overlaps with the region in which $2\,0\leftrightarrow1\,0$ conversion plays an important role in setting the relic abundance. For $m_{A'}=1~{\rm GeV}$, splittings around $\delta\simeq200~{\rm keV}$ therefore approximately mark the largest values for which the high-energy LZ recoil can remain compatible with conversion-driven freeze-out.

\section{Conclusion}
\label{sec:con}
%

In this work, we have explored the connection between conversion-driven freeze-out and the high-energy nuclear recoil recently reported by LZ within a singlet-doublet Dirac fermionic DM framework. We formulated a minimal framework for singlet-doublet Dirac fermionic DM and provided a possible UV completion generating the required singlet-doublet interaction. In the small-mixing regime, direct annihilation of the predominantly singlet-like $\chi_1$ becomes inefficient, while conversion processes transfer its population into the more strongly interacting doublet-like sector, which is subsequently depleted through
efficient electroweak annihilations. As the conversion rate becomes
inefficient, this transfer itself becomes the bottleneck controlling the
final $\chi_1$ abundance, leading to conversion-driven freeze-out, or coscattering. The small $\mathcal{O}(100\ {\rm keV})$ splitting between the two neutral states is essentially irrelevant for freeze-out, but becomes crucial for present-day endothermic scattering. Splittings in the range $\delta\simeq150$--$300~{\rm keV}$ allow the high-energy LZ recoil to be kinematically accessible while selecting the
high-velocity tail of the Galactic DM distribution. At the same time, the small singlet-doublet mixing required for conversion-driven freeze-out would ordinarily suppress direct detection. Introducing a light dark photon mediating the inelastic dark sector scattering provides the necessary enhancement of this transition while remaining compatible with existing searches for light vector mediators for the kinetic mixing considered here.
\vspace{1em}

The viable parameter space is determined by the interplay between physics at rather different scales. At lower DM masses, efficient SM-assisted electroweak conversion between the singlet- and doublet-like sectors drives the abundance below the observed value even as the dark gauge coupling is reduced, eventually requiring a smaller
singlet-doublet mixing that suppresses the LZ rate. Towards larger masses, an increasingly large dark gauge coupling is instead required to maintain efficient conversion, until the upper value adopted in
our scan is reached. For $m_{A'}=250~{\rm MeV}$, the light mediator provides sufficient enhancement for the LZ-sensitive bands corresponding to different values of $\delta$ to essentially cover the full relic-compatible region, including the deep conversion-driven parameter space characterised by the bottom range of the singlet-doublet mixing. Increasing $m_{A'}$ progressively suppresses the present-day scattering rate and moves the LZ sensitivity towards larger mixing angles. Consequently, for $m_{A'}=500~{\rm MeV}$ the overlap with the conversion-driven region is reduced, while for $m_{A'}=1~{\rm GeV}$ the small-mixing region is progressively lost and only the larger-mixing part remains accessible. The smaller dark coupling required in the latter case nevertheless extends the relic-compatible region to larger DM masses, with the combined requirements ultimately selecting DM masses of approximately $750$--$1600~{\rm GeV}$.

The crucial feature of this construction is that the connection between
freeze-out and direct detection is not merely one of complementary
constraints. In the coscattering regime, the relic abundance is governed
by the efficiency of the off-diagonal $\chi_1\leftrightarrow\chi_2$
conversion, making it intrinsically sensitive to the same dark coupling,
singlet-doublet mixing and mediator mass that determine the inelastic
transition probed by LZ. The light dark photon makes this otherwise weakly coupled regime experimentally accessible. The high-energy recoil can therefore test the interaction structure responsible for the non-standard thermal history of DM, allowing the same underlying physics to be probed during freeze-out in the early Universe and through scattering of Galactic DM today.

\section*{Acknowledgments}

SC acknowledges support from the UKRI Future Leader Fellowship DarkMAP (Ref. no. MR/Y034112/1) and thanks Rashidul Islam and Genevieve Belanger for useful discussions.

\appendix

\section{ UV completion of the singlet-doublet mixing operator}
\label{app:uv}
The dimension-5 interaction introduced in Sec.~\ref{sec:model},

\begin{align}
\mathcal L_{\rm EFT}\supset
-\left[
\frac{c}{\Lambda}\Phi\,\overline \Psi_L\widetilde H \xi_R+
\frac{c}{\Lambda}\Phi\,\overline \Psi_R\widetilde H \xi_L+
{\rm h.c.}
\right],
\label{eq:app1}
\end{align}

admits a simple renormalisable UV-completion through a heavy vector-like fermion $N$~\footnote{Here the EFT expansion refers only to the higher-dimensional singlet-doublet mixing operator, while the dark photon and dark Higgs are retained as explicit light degrees of freedom.}. We take

\begin{align}
N_{L,R}\sim(\mathbf 1,0)_{+1}
\label{}
\end{align}

under \(SU(2)_L\times U(1)_Y\times U(1)_D\). An additional \(Z_2\) symmetry is imposed under which $\xi$, $\Psi$, and $N$ are odd, while the SM fields, $\Phi$, and the dark gauge boson are even. The lightest state in the \(Z_2\)-odd sector is therefore stable.
The relevant renormalisable interactions can be written as

\begin{align}
\begin{aligned}
\mathcal L_{\rm UV}\supset{}&
-M_N\overline N N
\\
&-\left[
y_D\,\overline \Psi_L\widetilde H N_R
+y_S\,\Phi\,\overline N_L \xi_R
+y_D'\,\overline \Psi_R\widetilde H N_L
+y_S'\,\Phi\,\overline N_R \xi_L
+{\rm h.c.}
\right].
\label{eq:app3}
\end{aligned}
\end{align}

All interactions in Eq.~\ref{eq:app3} are invariant under the SM gauge group, \(U(1)_D\), and the imposed \(Z_2\). In particular,
\[
q_D(\overline \Psi)=-1,\qquad
q_D(N)=+1,
\]
for the first and third interactions, while
\[
q_D(\Phi)=+1,\qquad
q_D(\overline N)=-1
\]
for the second and fourth. At energies well below \(M_N\), integrating out \(N\) at tree level gives:

\begin{align}
\mathcal L_{\rm eff}\supset
-\left[
\frac{y_Dy_S}{M_N}
\Phi\,\overline \Psi_L\widetilde H \xi_R+
\frac{y_D'y_S'}{M_N}
\Phi\,\overline \Psi_R\widetilde H \xi_L+
{\rm h.c.}
\right]
+\mathcal O(M_N^{-2}),
\label{eq:app4}
\end{align}

and hence the matching conditions

\begin{align}
\boxed{
\frac{c_R}{\Lambda}=\frac{y_Dy_S}{M_N},
\qquad
\frac{c_L}{\Lambda}=\frac{y_D'y_S'}{M_N}.
}
\label{}
\end{align}
For simplicity, we may choose
\begin{align}
y_D=y_D',\qquad y_S=y_S',
\label{}
\end{align}
which gives
\begin{align}
c_L=c_R\equiv c,\qquad
\frac{c}{\Lambda}=\frac{y_Dy_S}{M_N}.
\label{}
\end{align}
After symmetry breaking,
\begin{align}
H=
\frac{1}{\sqrt2}
\begin{pmatrix}0\\v+h\end{pmatrix},
\qquad
\widetilde H=
\frac{1}{\sqrt2}
\begin{pmatrix}v+h\\0\end{pmatrix},
\qquad
\Phi=\frac{v_D+h_D}{\sqrt2},
\label{}
\end{align}
Eq.~\ref{eq:app1} becomes
\begin{align}
\mathcal L_{\rm EFT}\supset
-m
\left(1+\frac{h}{v}+\frac{h_D}{v_D}
+\frac{hh_D}{vv_D}\right)
(\overline \psi^0 \xi+\overline \xi \psi^0),
\qquad
m=\frac{cvv_D}{2\Lambda}.
\label{}
\end{align}
Using
\begin{align}
\xi=c_\theta\chi_1+s_\theta\chi_2,
\qquad
\psi^0=-s_\theta\chi_1+c_\theta\chi_2,
\label{}
\end{align}
one obtains
\begin{align}
\begin{aligned}
\overline \psi^0\xi+\overline \xi \psi^0
={}&-2s_\theta c_\theta
\left(\overline{\chi_1}\chi_1-\overline\chi_2\chi_2\right)
\\
&+\left(c_\theta^2-s_\theta^2\right)
(\overline{\chi_1}\chi_2+\overline{\chi_2}\chi_1),
\end{aligned}
\label{}
\end{align}
which reproduces the scalar interactions used in Sec.~\ref{sec:model}. In particular,
\begin{align}
m=(m_{\chi_2}-m_{\chi_1})\,s_\theta c_\theta
=\delta s_\theta c_\theta .
\label{}
\end{align}
The heavy state $N$ therefore provides a renormalisable realisation of the effective mixing operator without introducing additional light degrees of freedom. Since all new fermions are vector-like under the gauge symmetries, their
contributions to the gauge anomalies cancel between the corresponding
left- and right-handed components.

\section{Momentum dependence of the coscattering rate}
\label{app:kinetic_equilibrium}

The momentum-integrated Boltzmann equations employed in the main text assume that departures of the DM phase space distribution from its equilibrium shape do not significantly affect the relic density calculation. This assumption can become non-trivial in some conversion-driven freezeout scenarios, where an endothermic conversion process may preferentially deplete the high-momentum part of the DM distribution. Different momentum modes can then decouple at different temperatures, requiring a momentum-dependent treatment of the Boltzmann equation.

We explicitly test this possibility for the coscattering processes relevant in our model. In particular, we consider the dark bath conversions

\begin{align}
    \chi_1 A' &\leftrightarrow \chi_2 h_D ,
    \\
    \chi_1 h_D &\leftrightarrow \chi_2 A' ,
    \label{eq:dark_conversion_app}
\end{align}

which belong to the $1\,0\leftrightarrow2\,0$ reaction class.
For a fixed incoming DM momentum, it is convenient to introduce

\begin{align}
    q \equiv \frac{|\mathbf p_{\chi_1}|}
                   {\sqrt{m_{\chi_1}T}},
\end{align}

such that $q=\mathcal O(1)$ corresponds to the thermally populated non-relativistic part of the DM distribution. 

For a generic conversion

\begin{align}
    \chi_1(p_1)+a(p_2)\to\chi_2(p_4)+b(p_3),
\end{align}

where $a=A',h_D$, the interaction rate of a DM particle with fixed momentum $\mathbf p_1$ is~\cite{Belanger:2020npe}

\begin{align}
\Gamma_{\chi_1 a\to\chi_2 b}(q,T)
={\frac{g_a}{(2\pi)^2}}
\int_0^\infty dp_2\,p_2^2 f_a(E_2,T)
\int_{-1}^{1}d\cos\vartheta\,
v_{\rm M\o l}\,
\sigma_{\chi_1a\to\chi_2b}(s),
\label{eq:momentum_rate}
\end{align}
where
\begin{align}
v_{\rm M\o l}
&=
\frac{\sqrt{(p_1\cdot p_2)^2-m_{\chi_1}^2m_a^2}}{E_1E_2},
\\
s&=m_{\chi_1}^2+m_a^2+
2\left(E_1E_2-|\mathbf p_1||\mathbf p_2|
\cos\vartheta\right),
\end{align}
with $|\mathbf p_1|=q\sqrt{m_{\chi_1}T}$. The cross section is
evaluated from the full tree-level matrix element,
\begin{align}
\sigma(s)=
{1\over16\pi\lambda(s,m_{\chi_1}^2,m_a^2)}
\int_{t_-}^{t_+}dt\,
\overline{|\mathcal M(s,t)|^2},
\end{align}
where $\lambda$ denotes the K\"all\'en function. We use the equilibrium distribution $f_a(E_2,T)$ for the dark bath particle.

The evolution of the most conversion-dominated viable points in our scans in Sec.~\ref{sec:bound} shows that, for all practical purposes, the DM abundance begins to depart appreciably from equilibrium around
\begin{align}
    x_{\rm fo}\equiv {m_{\chi_1}\over T_{\rm fo}}\simeq25.
\end{align}
We therefore evaluate the momentum dependence at $x=25$ and define
\begin{align}
    \kappa(q,T)
    \equiv
    {\Gamma_{\rm conv}(q,T)
     \over
     \Gamma_{\rm conv}(0,T)}.
    \label{eq:Rmomentum}
\end{align}
\begin{figure}[htb]
    \centering
    \includegraphics[width=0.48\textwidth]{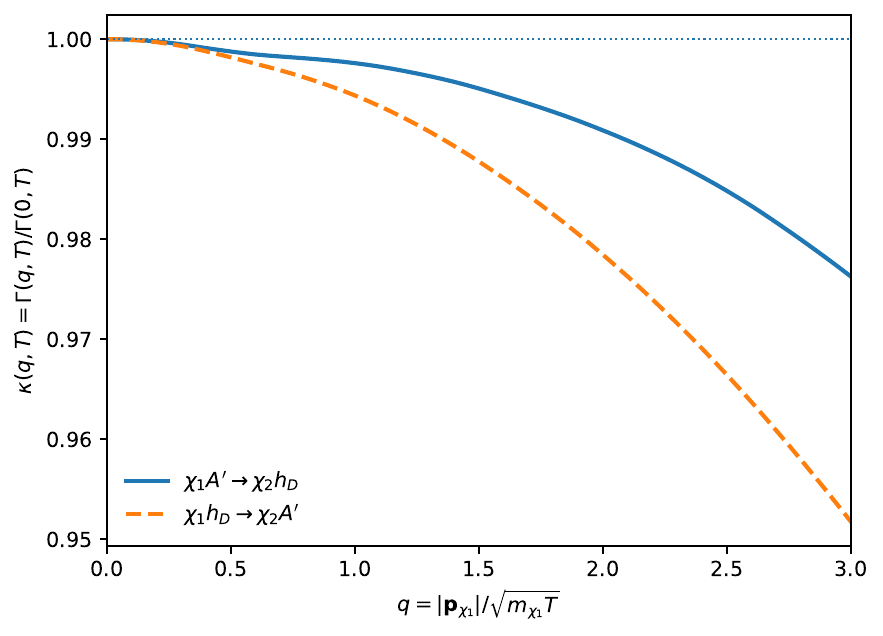}
    \hfill
    \includegraphics[width=0.48\textwidth]{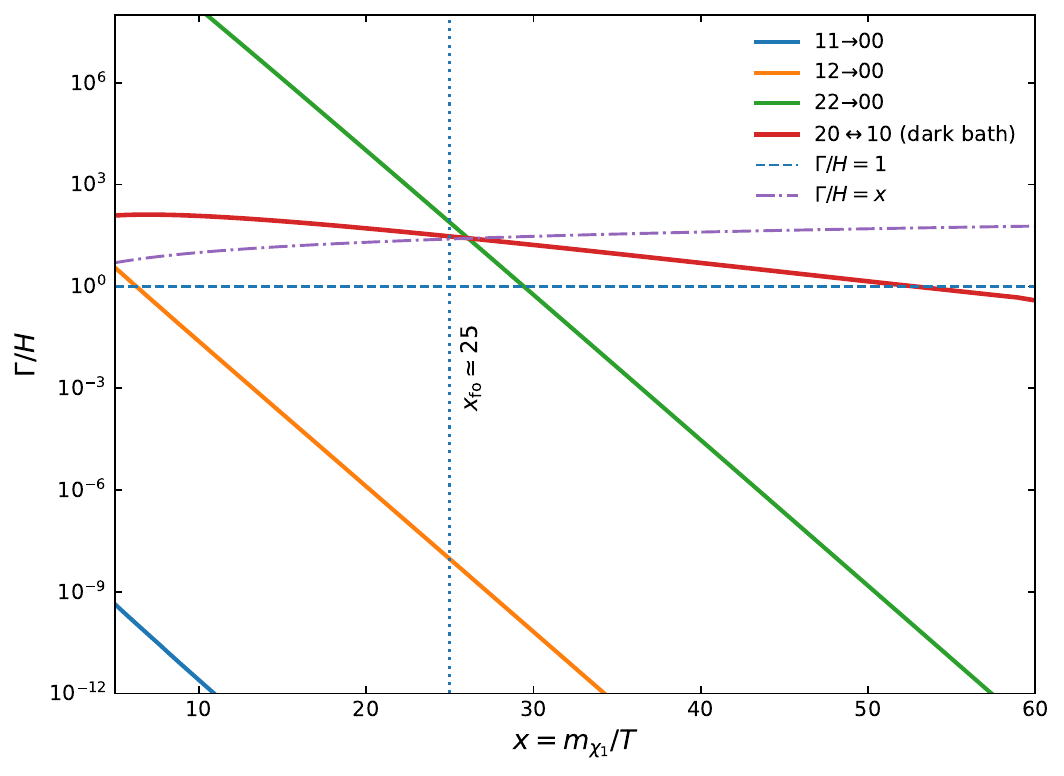}
    \caption{Momentum dependence of the dominant dark bath conversion processes
and comparison of the relevant interaction rates with the Hubble expansion
rate. {\it left}: ratio $\kappa(q,T)=\Gamma(q,T)/\Gamma(0,T)$ for
$\chi_1A'\to\chi_2h_D$ and $\chi_1h_D\to\chi_2A'$, showing only a mild
momentum dependence over the thermally relevant range. {\it right}: interaction
rates normalised to the Hubble rate as a function of
$x=m_{\chi_1}/T$. The different curves show the annihilation,
coannihilation and conversion classes indicated in the legend, with the
dark bath contribution to $20\leftrightarrow10$ shown separately. The
horizontal line denotes $\Gamma/H=1$, while the dot-dashed curve shows the
stronger $\Gamma/H=x$ criterion as a reference for maintaining chemical
equilibrium; neither is used as a definition of the relic freeze-out
point. The vertical dotted line indicates $x\simeq25$, around which the
total odd-sector abundance begins to depart from equilibrium.}
\label{fig:app}
\end{figure}

As shown in the {\it left} panel of Fig.~\ref{fig:app}, the momentum dependence remains mild throughout the thermally populated region. At $x=25$, even for $q=3$ we find

\begin{align}
\kappa(3,T)\simeq0.92
\qquad &\text{for}\qquad
\chi_1A'\to\chi_2h_D,
\\
\kappa(3,T)\simeq0.90
\qquad &\text{for}\qquad
\chi_1h_D\to\chi_2A'.
\end{align}

Thus the rates of the dominant dark bath conversion processes vary by at
most approximately $10\%$ over $q\lesssim3$ at the epoch relevant for
freeze-out. We therefore find no evidence for the strongly momentum-selective decoupling characteristic of more extreme coscattering scenarios.

As mentioned in the main text, it is important to emphasise that freeze-out of the total DM abundance does not require the different dark sector states to cease interacting efficiently with one another. This is illustrated in the right panel of Fig.~\ref{fig:app}, where we show the ratios $\Gamma/H$ for the relevant reaction classes at a representative relic density benchmark. The combined odd sector abundance begins to depart from equilibrium with the thermal bath around $x=m_{\chi_1}/T\simeq25$. At this epoch, the $2\,2\leftrightarrow0\,0$ reactions remain very efficient and provide the dominant number-depleting sink for the heavier sector, while the dark bath $2\,0\leftrightarrow1\,0$ conversions continue to
transfer particles between $\chi_1$ and $\chi_2$. In contrast, direct $\chi_1$ annihilation ($1\,1\leftrightarrow0\,0$) and co-annihilation
($1\,2\leftrightarrow0\,0$) have already become negligible.

The subsequent evolution therefore reflects a conversion-driven depletion chain: the abundance of the predominantly singlet-like $\chi_1$ state is depleted through conversion into the heavier sector,
which in turn remains efficiently depleted through
$2\,2\leftrightarrow0\,0$ reactions. The $2\,2\leftrightarrow0\,0$ rate decreases rapidly relative to the Hubble rate as the heavier states become Boltzmann-suppressed, whereas the dark bath conversion rate decreases more slowly and continues to maintain efficient communication between the two odd sectors. Consequently, internal dark sector equilibration can persist well beyond the onset of freeze-out of the total abundance. Across the representative benchmark points considered here, this  transition from the onset of departure of the total abundance from equilibrium to the eventual loss of efficient internal conversion occurs over approximately $25\lesssim x\lesssim60$.

\bibliographystyle{JHEP}
\bibliography{ref}

@article{Bertone:2004pz,
    author = "Bertone, Gianfranco and Hooper, Dan and Silk, Joseph",
    title = "{Particle dark matter: Evidence, candidates and constraints}",
    eprint = "hep-ph/0404175",
    archivePrefix = "arXiv",
    reportNumber = "FERMILAB-PUB-04-047-A",
    doi = "10.1016/j.physrep.2004.08.031",
    journal = "Phys. Rept.",
    volume = "405",
    pages = "279--390",
    year = "2005"
}

@article{Arcadi:2017kky,
    author = "Arcadi, Giorgio and Dutra, Ma{\'\i}ra and Ghosh, Pradipta and Lindner, Manfred and Mambrini, Yann and Pierre, Mathias and Profumo, Stefano and Queiroz, Farinaldo S.",
    title = "{The waning of the WIMP? A review of models, searches, and constraints}",
    eprint = "1703.07364",
    archivePrefix = "arXiv",
    primaryClass = "hep-ph",
    doi = "10.1140/epjc/s10052-018-5662-y",
    journal = "Eur. Phys. J. C",
    volume = "78",
    number = "3",
    pages = "203",
    year = "2018"
}

@article{Arcadi:2024ukq,
    author = "Arcadi, Giorgio and Cabo-Almeida, David and Dutra, Ma{\'\i}ra and Ghosh, Pradipta and Lindner, Manfred and Mambrini, Yann and Neto, Jacinto P. and Pierre, Mathias and Profumo, Stefano and Queiroz, Farinaldo S.",
    title = "{The Waning of the WIMP: Endgame?}",
    eprint = "2403.15860",
    archivePrefix = "arXiv",
    primaryClass = "hep-ph",
    doi = "10.1140/epjc/s10052-024-13672-y",
    journal = "Eur. Phys. J. C",
    volume = "85",
    number = "2",
    pages = "152",
    year = "2025"
}

@article{Liu:2017drf,
    author = "Liu, Jianglai and Chen, Xun and Ji, Xiangdong",
    title = "{Current status of direct dark matter detection experiments}",
    eprint = "1709.00688",
    archivePrefix = "arXiv",
    primaryClass = "astro-ph.CO",
    doi = "10.1038/nphys4039",
    journal = "Nature Phys.",
    volume = "13",
    number = "3",
    pages = "212--216",
    year = "2017"
}

@article{Vahsen:2021gnb,
    author = "Vahsen, Sven E. and O'Hare, Ciaran A. J. and Loomba, Dinesh",
    title = "{Directional Recoil Detection}",
    eprint = "2102.04596",
    archivePrefix = "arXiv",
    primaryClass = "physics.ins-det",
    doi = "10.1146/annurev-nucl-020821-035016",
    journal = "Ann. Rev. Nucl. Part. Sci.",
    volume = "71",
    pages = "189--224",
    year = "2021"
}

@article{LZ:2026axp,
    author = "Akerib, D. S. and others",
    collaboration = "LZ",
    title = "Search for dark matter particle interactions in an extended nuclear recoil energy window with the LUX-ZEPLIN (LZ) experiment",
    eprint = "2609.02823",
    archivePrefix = "arXiv",
    primaryClass = "hep-ex",
    doi = "10.17182/hepdata.182472.v1",
    month = "9",
    year = "2026"
}

@article{Fan:2026kxx,
  author = {Fan, JiJi and Reece, Matthew},
  title = {{Higgsino Above the Sea of Fog}},
  eprint = {2609.01504},
  archiveprefix = {arXiv},
  primaryclass = {hep-ph},
  month = {9},
  year = {2026},
}

@article{Wu:2026nhi,
  author = {Wu, Lei and Zhang, Yang and Zhu, Bin},
  title = {{TeV Higgsino Dark Matter from LZ Nuclear Recoil to Fermi-LAT Gamma Rays}},
  eprint = {2609.01590},
  archiveprefix = {arXiv},
  primaryclass = {hep-ph},
  month = {9},
  year = {2026},
}

@article{Freese:2026sga,
    author = "Freese, Katherine and Theodosopoulos, Dionysios P.",
    title = "{Higgsino Dark Matter Interpretation of the LUX-ZEPLIN 248 keV Nuclear-Recoil Event}",
    eprint = "2609.01583",
    archivePrefix = "arXiv",
    primaryClass = "hep-ph",
    month = "9",
    year = "2026"
}

@article{Du:2026guj,
    author = "Du, Xiaokang and Wang, Fei",
    title = "{TeV Higgsino Interpretation of the LZ High-Recoil Event with Intermediate-Scale Electroweak Gauginos}",
    eprint = "2609.04163",
    archivePrefix = "arXiv",
    primaryClass = "hep-ph",
    month = "9",
    year = "2026"
}

@article{Langhoff:2026ujr,
    author = "Langhoff, Kevin",
    title = "{Heavy Higgsino Interpretation of the LZ Event}",
    eprint = "2609.09385",
    archivePrefix = "arXiv",
    primaryClass = "hep-ph",
    month = "9",
    year = "2026"
}

@article{Bisal:2026khf,
    author = "Bisal, Subhadip and Cao, Junjie and Li, Fei",
    title = "{Higgsino Dark Matter Interpretation of the LZ High-Recoil Event in the GNMSSM with TeV-Scale Gauginos}",
    eprint = "2609.07811",
    archivePrefix = "arXiv",
    primaryClass = "hep-ph",
    month = "9",
    year = "2026"
}

@article{Frolovsky:2026tvq,
    author = "Frolovsky, Daniel and Ketov, Sergei V.",
    title = "{Higgsino dark matter in the Starobinsky supergravity with the MSSM in light of the LUX-ZEPLIN event}",
    eprint = "2609.11241",
    archivePrefix = "arXiv",
    primaryClass = "hep-ph",
    reportNumber = "IPMU26-0034",
    month = "9",
    year = "2026"
}

@article{Yin:2026jnn,
  author = {Yin, Wen},
  title = {{A PQ-Symmetric High-Scale SUSY Interpretation of the LZ High-Energy Recoil}},
  eprint = {2609.01892},
  archiveprefix = {arXiv},
  primaryclass = {hep-ph},
  month = {9},
  year = {2026},
}

@article{Smirnov:2026aqk,
    author = "Smirnov, Juri and Griffith, Spencer and Beacom, John F.",
    title = "{Inelastic Signatures of Electroweak Dark Matter}",
    eprint = "2609.04144",
    archivePrefix = "arXiv",
    primaryClass = "hep-ph",
    month = "9",
    year = "2026"
}

@article{Nomura:2026qyq,
    author = "Nomura, Yasunori",
    title = "{Dark Matter as the Z{\_}2 Partner of the Standard Model Higgs Boson}",
    eprint = "2609.02505",
    archivePrefix = "arXiv",
    primaryClass = "hep-ph",
    reportNumber = "RIKEN-iTHEMS-Report-26",
    month = "9",
    year = "2026"
}

@article{Lou:2026idn,
    author = "Lou, Yuanchao and Lu, Chih-Ting",
    title = "{Fermionic Dark Matter Absorption and the High-Energy Event in LUX-ZEPLIN}",
    eprint = "2609.01592",
    archivePrefix = "arXiv",
    primaryClass = "hep-ph",
    month = "9",
    year = "2026"
}

@article{Su:2026rwz,
    author = "Su, Liangliang and Yang, Jin Min and Yang, Wen-Na",
    title = "{Inelastic Dark Matter Signature at High Recoil Energy in LUX-ZEPLIN and CRESST}",
    eprint = "2609.01475",
    archivePrefix = "arXiv",
    primaryClass = "hep-ph",
    month = "9",
    year = "2026"
}

@article{DiMauro:2026ldr,
    author = "Di Mauro, Mattia",
    title = "{Dark Matter at the Kinematic Edge: Interpreting the 248 keV LZ Nuclear-Recoil Candidate}",
    eprint = "2609.02608",
    archivePrefix = "arXiv",
    primaryClass = "hep-ph",
    month = "9",
    year = "2026"
}

@article{Yamashita:2026ump,
    author = "Yamashita, Kimiko",
    title = "{Inelastic Dark Photon Dark Matter for the LUX-ZEPLIN High-Recoil Event and the Galactic Halo Gamma-Ray Excess}",
    eprint = "2609.02868",
    archivePrefix = "arXiv",
    primaryClass = "hep-ph",
    month = "9",
    year = "2026"
}

@article{Chattopadhyay:2026ryw,
    author = "Chattopadhyay, Utpal and Das, Debottam and Puri, Rahul and Roy, Joydeep",
    title = "{Sub-TeV Singlino Dark Matter in light from Sagittarius A$^\ast$ and LUX-ZEPLIN Nuclear-Recoil Event}",
    eprint = "2609.02994",
    archivePrefix = "arXiv",
    primaryClass = "hep-ph",
    month = "9",
    year = "2026"
}

@article{deLima:2026shq,
    author = "de Lima, Carlos Henrique",
    title = "{Exothermic Dark Matter at LZ}",
    eprint = "2609.05204",
    archivePrefix = "arXiv",
    primaryClass = "hep-ph",
    month = "9",
    year = "2026"
}

@article{Wang:2026ytg,
    author = "Wang, Lei and Xiao, Yang",
    title = "{The Inert Doublet Model of Dark Matter and the LUX-ZEPLIN High-Recoil Event}",
    eprint = "2609.06571",
    archivePrefix = "arXiv",
    primaryClass = "hep-ph",
    month = "9",
    year = "2026"
}

@article{Dent:2026bji,
    author = "Dent, James B. and Newstead, Jayden L.",
    title = "{Exothermic and Endothermic Inelastic Dark Matter Interpretations at LZ: Sideband Constraints and Future Prospects}",
    eprint = "2609.04673",
    archivePrefix = "arXiv",
    primaryClass = "hep-ph",
    month = "9",
    year = "2026"
}

@article{McCabe:2026crm,
    author = "McCabe, Christopher",
    title = "{Seasonal dark matter from the LUX-ZEPLIN high-energy event}",
    eprint = "2609.04181",
    archivePrefix = "arXiv",
    primaryClass = "hep-ph",
    month = "9",
    year = "2026"
}

@article{Okada:2026upm,
    author = "Okada, Hiroshi and Shigekami, Yoshihiro and Wu, Jia-Jun",
    title = "{Can a minimal radiative seesaw explain the LZ 248 keV event?}",
    eprint = "2609.13038",
    archivePrefix = "arXiv",
    primaryClass = "hep-ph",
    month = "9",
    year = "2026"
}

@article{Ahmed:2026qjg,
    author = "Ahmed, Waqas and Leontaris, George K.",
    title = "{A Dark-Dimension Origin of Geometric Inelastic Dark Matter: The LUX-ZEPLIN High-Recoil Event and Multi-Target Tests}",
    eprint = "2609.07138",
    archivePrefix = "arXiv",
    primaryClass = "hep-ph",
    month = "9",
    year = "2026"
}

@article{Murayama:2026apt,
    author = "Murayama, Hitoshi and Noether, Bea",
    title = "{Heavy Inert Doublet Reconciles LZ and IceCube}",
    eprint = "2609.28819",
    archivePrefix = "arXiv",
    primaryClass = "hep-ph",
    month = "9",
    year = "2026"
}

@article{Gemmell:2026yaw,
    author = "Gemmell, Caleb and Hooper, Dan and Krnjaic, Gordan",
    title = "{A Simple Dark Matter Model to Explain the LZ Event and Galactic Center Excess}",
    eprint = "2609.26570",
    archivePrefix = "arXiv",
    primaryClass = "hep-ph",
    reportNumber = "FERMILAB-PUB-26-0702-T",
    month = "9",
    year = "2026"
}

@article{Okada:2026eol,
    author = "Okada, Nobuchika and Seto, Osamu",
    title = "{Inelastic $B-L$ scalar dark matter and the LUX-ZEPLIN event}",
    eprint = "2609.06909",
    archivePrefix = "arXiv",
    primaryClass = "hep-ph",
    reportNumber = "EPHOU-26-011",
    month = "9",
    year = "2026"
}

@article{Das:2026buc,
    author = "Das, Arindam and Nomura, Takaaki",
    title = "{Effect of inelastic scalar dark matter in hidden $U(1)$ scenario after the LZ nuclear recoil}",
    eprint = "2609.15600",
    archivePrefix = "arXiv",
    primaryClass = "hep-ph",
    month = "9",
    year = "2026"
}

@article{De:2026win,
    author = "De, Bibhabasu",
    title = "{The 248 keV LZ Recoil: A Possible Hint of Non-SM-Like Quark Yukawa Couplings with a Scalar-Portal Dark Matter}",
    eprint = "2609.23096",
    archivePrefix = "arXiv",
    primaryClass = "hep-ph",
    month = "9",
    year = "2026"
}

@article{Visinelli:2026kgt,
    author = "Visinelli, Luca",
    title = "{A Peccei-Quinn Origin for Inelastic Electroweak Dark Matter after LUX-ZEPLIN}",
    eprint = "2609.02807",
    archivePrefix = "arXiv",
    primaryClass = "hep-ph",
    month = "9",
    year = "2026"
}

@article{An:2026pkc,
    author = "An, Haipeng and Gao, Fei and Liu, Jia and Liu, Minghao and Xu, Changlong",
    title = "{Cosmological Constrained Axion-Portal Inelastic Dark Matter for the LZ Event}",
    eprint = "2609.17412",
    archivePrefix = "arXiv",
    primaryClass = "hep-ph",
    month = "9",
    year = "2026"
}

@article{Unwin:2026rdp,
    author = "Unwin, James",
    title = "{Axion Portal Dark Matter and the LUX-ZEPLIN High-Recoil Event}",
    eprint = "2609.04186",
    archivePrefix = "arXiv",
    primaryClass = "hep-ph",
    month = "9",
    year = "2026"
}

@article{Aghaie:2026vsu,
    author = "Aghaie, Mohammad and Strumia, Alessandro",
    title = "{Neutron disappearance and the LZ nuclear recoil event}",
    eprint = "2609.09037",
    archivePrefix = "arXiv",
    primaryClass = "hep-ph",
    month = "9",
    year = "2026"
}

@article{Bandyopadhyay:2026gjw,
    author = "Bandyopadhyay, Disha and Borah, Debasish and Borah, Pankaj",
    title = "{LZ nuclear recoil event from inelastic singlet-doublet scalar dark matter}",
    eprint = "2609.07451",
    archivePrefix = "arXiv",
    primaryClass = "hep-ph",
    month = "9",
    year = "2026"
}

@article{Borah:2026zwf,
    author = "Borah, Debasish and Sahoo, Sujit Kumar and Sahu, Narendra and Sharma, Shashwat",
    title = "{Inelastic Singlet-Doublet Fermion Dark Matter in light of the 248 keV LZ event}",
    eprint = "2609.07800",
    archivePrefix = "arXiv",
    primaryClass = "hep-ph",
    month = "9",
    year = "2026"
}

@article{Brdar:2026ukx,
    author = "Brdar, Vedran and Chattopadhyay, Dibya S.",
    title = "{A Narrow Neutrino Window for the LZ Event}",
    eprint = "2609.30255",
    archivePrefix = "arXiv",
    primaryClass = "hep-ph",
    month = "9",
    year = "2026"
}

@article{Jeesun:2026vzo,
    author = "Jeesun, Sk and Majumdar, Anirban",
    title = "{Atmospheric neutrino up-scattering explanation of LZ 2026 excess}",
    eprint = "2609.04185",
    archivePrefix = "arXiv",
    primaryClass = "hep-ph",
    month = "9",
    year = "2026"
}

@article{Chattaraj:2026fxn,
    author = "Chattaraj, Ayan and Majumdar, Anirban and Papoulias, Dimitrios K. and Srivastava, Rahul",
    title = "{Can Elastic Neutrino Scattering Account for the LZ230616 Event?}",
    eprint = "2609.10504",
    archivePrefix = "arXiv",
    primaryClass = "hep-ph",
    month = "9",
    year = "2026"
}

@article{Pospelov:2026ewn,
    author = "Pospelov, Maxim and Ramani, Harikrishnan",
    title = "{Strong Constraints on Higgsino Dark Matter from Solar Capture}",
    eprint = "2609.02775",
    archivePrefix = "arXiv",
    primaryClass = "hep-ph",
    month = "9",
    year = "2026"
}

@article{Bose:2026ndd,
    author = "Bose, Debajit and others",
    title = "{Not so good $\nu$s for Higgsino dark matter as LZ excess: stringent limits from Super-Kamiokande and IceCube}",
    eprint = "2609.07807",
    archivePrefix = "arXiv",
    primaryClass = "hep-ph",
    month = "9",
    year = "2026"
}

@article{Nguyen:2026lui,
    author = "Nguyen, Thong T. Q. and Linden, Tim and Hooper, Dan",
    title = "{Solar Neutrino Constraints on Inelastic Dark Matter Scattering in Light of Recent LUX-ZEPLIN Observations}",
    eprint = "2609.11833",
    archivePrefix = "arXiv",
    primaryClass = "hep-ph",
    month = "9",
    year = "2026"
}

@article{DiMauro:2026dqp,
    author = "Di Mauro, Mattia and Shaikh, Halim",
    title = "{Solar Capture Tests of Inelastic Dark Matter after the LZ High-Recoil Event}",
    eprint = "2609.06760",
    archivePrefix = "arXiv",
    primaryClass = "hep-ph",
    month = "9",
    year = "2026"
}

@article{Rodd:2026tyn,
    author = "Rodd, Nicholas L. and Safdi, Benjamin R. and Slatyer, Tracy R. and Xu, Weishuang Linda",
    title = "{Confronting the Higgsino Interpretation of the LZ Event with the High-Energy Sideband}",
    eprint = "2609.04175",
    archivePrefix = "arXiv",
    primaryClass = "hep-ph",
    month = "9",
    year = "2026"
}

@article{Paul:2024prs,
    author = "Paul, Partha Kumar and Sahoo, Sujit Kumar and Sahu, Narendra",
    title = "{Anatomy of singlet-doublet dark matter relic: annihilation, co-annihilation, co-scattering, and freeze-in}",
    eprint = "2412.02607",
    archivePrefix = "arXiv",
    primaryClass = "hep-ph",
    doi = "10.1088/1475-7516/2025/10/053",
    journal = "JCAP",
    volume = "10",
    pages = "053",
    year = "2025",
    note = "[Erratum: JCAP 01, E02 (2026)]"
}

@article{Bhattiprolu:2026hnp,
    author = "Bhattiprolu, Prudhvi N. and Pierce, Aaron and Takla, Alexander",
    title = "{Singlet-doublet dark matter beyond freeze-out}",
    eprint = "2608.12480",
    archivePrefix = "arXiv",
    primaryClass = "hep-ph",
    reportNumber = "LITP 26-15",
    month = "8",
    year = "2026"
}

@article{Belanger:2026asz,
    author = "Belanger, G. and Belyaev, A. and Bernal, N. and Boudjema, F. and Chakraborti, S. and Goudelis, A. and Pukhov, A.",
    title = "{micrOMEGAs 7: Beyond standard cosmology}",
    eprint = "2606.06645",
    archivePrefix = "arXiv",
    primaryClass = "hep-ph",
    reportNumber = "IPPP/26/46",
    month = "6",
    year = "2026"
}

@misc{Cyncynates:lzmap,
    author = {David Cyncynates},
    title  = {LUX-ZEPLIN 248 keV Paper Map: High-Recoil Literature},
    year={2026},
    howpublished = {\url{https://davidcyncynates.github.io/lz-paper-map/}}
}

@article{XENON:2022avm,
    author = "Aprile, E. and others",
    collaboration = "XENON",
    title = "{Effective field theory and inelastic dark matter results from XENON1T}",
    eprint = "2210.07591",
    archivePrefix = "arXiv",
    primaryClass = "hep-ex",
    doi = "10.1103/PhysRevD.109.112017",
    journal = "Phys. Rev. D",
    volume = "109",
    number = "11",
    pages = "112017",
    year = "2024"
}

@article{Cooley:2021rws,
    author = "Cooley, Jodi",
    title = "{Dark Matter direct detection of classical WIMPs}",
    eprint = "2110.02359",
    archivePrefix = "arXiv",
    primaryClass = "hep-ph",
    doi = "10.21468/SciPostPhysLectNotes.55",
    journal = "SciPost Phys. Lect. Notes",
    volume = "55",
    pages = "1",
    year = "2022"
}

@article{Vietze:2014vsa,
    author = "Vietze, L. and Klos, P. and Men{\'e}ndez, J. and Haxton, W. C. and Schwenk, A.",
    title = "{Nuclear structure aspects of spin-independent WIMP scattering off xenon}",
    eprint = "1412.6091",
    archivePrefix = "arXiv",
    primaryClass = "nucl-th",
    doi = "10.1103/PhysRevD.91.043520",
    journal = "Phys. Rev. D",
    volume = "91",
    number = "4",
    pages = "043520",
    year = "2015"
}

@article{ENGEL1991114,
title = {Nuclear form factors for the scattering of weakly interacting massive particles},
journal = {Physics Letters B},
volume = {264},
number = {1},
pages = {114-119},
year = {1991},
issn = {0370-2693},
doi = {https://doi.org/10.1016/0370-2693(91)90712-Y},
url = {https://www.sciencedirect.com/science/article/pii/037026939190712Y},
author = {J. Engel}
}

@article{BaBar:2014zli,
    author = "Lees, J. P. and others",
    collaboration = "BaBar",
    title = "{Search for a Dark Photon in $e^+e^-$ Collisions at BaBar}",
    eprint = "1406.2980",
    archivePrefix = "arXiv",
    primaryClass = "hep-ex",
    reportNumber = "BABAR-PUB-14-002, SLAC-PUB-15979",
    doi = "10.1103/PhysRevLett.113.201801",
    journal = "Phys. Rev. Lett.",
    volume = "113",
    number = "20",
    pages = "201801",
    year = "2014"
}

@article{LHCb:2017trq,
    author = "Aaij, Roel and others",
    collaboration = "LHCb",
    title = "{Search for Dark Photons Produced in 13 TeV $pp$ Collisions}",
    eprint = "1710.02867",
    archivePrefix = "arXiv",
    primaryClass = "hep-ex",
    reportNumber = "LHCB-PAPER-2017-038, CERN-EP-2017-248",
    doi = "10.1103/PhysRevLett.120.061801",
    journal = "Phys. Rev. Lett.",
    volume = "120",
    number = "6",
    pages = "061801",
    year = "2018"
}

@article{McKenna:2018xjb,
    author = "McKenna, Janis",
    editor = "Bar{\textasciitilde}ao, Fernando and others",
    collaboration = "BaBar",
    title = "{Search for Low-Mass Dark Sector New Physics States at BABAR}",
    doi = "10.1088/1742-6596/1137/1/012041",
    journal = "J. Phys. Conf. Ser.",
    volume = "1137",
    number = "1",
    pages = "012041",
    year = "2018"
}

@article{Baxter:2021pqo,
    author = "Baxter, D. and others",
    title = "{Recommended conventions for reporting results from direct dark matter searches}",
    eprint = "2105.00599",
    archivePrefix = "arXiv",
    primaryClass = "hep-ex",
    doi = "10.1140/epjc/s10052-021-09655-y",
    journal = "Eur. Phys. J. C",
    volume = "81",
    number = "10",
    pages = "907",
    year = "2021"
}

@article{LEWIN199687,
title = {Review of mathematics, numerical factors, and corrections for dark matter experiments based on elastic nuclear recoil},
journal = {Astroparticle Physics},
volume = {6},
number = {1},
pages = {87-112},
year = {1996},
issn = {0927-6505},
doi = {https://doi.org/10.1016/S0927-6505(96)00047-3},
url = {https://www.sciencedirect.com/science/article/pii/S0927650596000473},
author = {J.D. Lewin and P.F. Smith}
}

@article{Belanger:2020npe,
    author = "B{\'e}langer, Genevi{\`e}ve and Delaunay, C{\'e}dric and Pukhov, Alexander and Zaldivar, Bryan",
    title = "{Dark matter abundance from the sequential freeze-in mechanism}",
    eprint = "2005.06294",
    archivePrefix = "arXiv",
    primaryClass = "hep-ph",
    reportNumber = "LAPTH-021/20, IFT-UAM/CSIC-20-65",
    doi = "10.1103/PhysRevD.102.035017",
    journal = "Phys. Rev. D",
    volume = "102",
    number = "3",
    pages = "035017",
    year = "2020"
}

@article{Brummer:2019inq,
    author = {Br{\"u}mmer, F.},
    title = "{Coscattering in next-to-minimal dark matter and split supersymmetry}",
    eprint = "1910.01549",
    archivePrefix = "arXiv",
    primaryClass = "hep-ph",
    doi = "10.1007/JHEP01(2020)113",
    journal = "JHEP",
    volume = "01",
    pages = "113",
    year = "2020"
}

@article{Liu:2025swd,
    author = "Liu, Ang and Han, Zhi-Long and Huang, Fei and Shao, Feng-Lan and Wang, Wei",
    title = "{Coscattering dark matter in the inverse scotogenic model}",
    eprint = "2510.13231",
    archivePrefix = "arXiv",
    primaryClass = "hep-ph",
    doi = "10.1007/JHEP06(2026)218",
    journal = "JHEP",
    volume = "06",
    pages = "218",
    year = "2026"
}

@article{Cheng:2018vaj,
    author = "Cheng, Hsin-Chia and Li, Lingfeng and Zheng, Rui",
    title = "{Coscattering/Coannihilation Dark Matter in a Fraternal Twin Higgs Model}",
    eprint = "1805.12139",
    archivePrefix = "arXiv",
    primaryClass = "hep-ph",
    doi = "10.1007/JHEP09(2018)098",
    journal = "JHEP",
    volume = "09",
    pages = "098",
    year = "2018"
}

@article{DiazSaez:2024nrq,
    author = {D{\'\i}az S{\'a}ez, Basti{\'a}n and Lahiri, Jayita and M{\"o}hling, Kilian},
    title = "{Coscattering in the extended singlet-scalar Higgs portal}",
    eprint = "2404.19057",
    archivePrefix = "arXiv",
    primaryClass = "hep-ph",
    doi = "10.1088/1475-7516/2024/10/001",
    journal = "JCAP",
    volume = "10",
    pages = "001",
    year = "2024"
}

@article{Belanger:2025wjh,
    author = "B{\'e}langer, Genevi{\`e}ve and Mitra, Manimala and Padhan, Rojalin and Roy, Abhishek",
    title = "{Illuminating scalar dark matter co-scattering in EFT with monophoton signatures}",
    eprint = "2508.06040",
    archivePrefix = "arXiv",
    primaryClass = "hep-ph",
    reportNumber = "CQUeST-2025-0756, IP/BBSR/2025-02",
    doi = "10.1007/JHEP01(2026)007",
    journal = "JHEP",
    volume = "01",
    pages = "007",
    year = "2026"
}

@article{Garny:2017rxs,
    author = {Garny, Mathias and Heisig, Jan and L{\"u}lf, Benedikt and Vogl, Stefan},
    title = "{Coannihilation without chemical equilibrium}",
    eprint = "1705.09292",
    archivePrefix = "arXiv",
    primaryClass = "hep-ph",
    reportNumber = "TUM-HEP-1085-17, TTK-17-18",
    doi = "10.1103/PhysRevD.96.103521",
    journal = "Phys. Rev. D",
    volume = "96",
    number = "10",
    pages = "103521",
    year = "2017"
}

@article{DAgnolo:2017dbv,
    author = "D'Agnolo, Raffaele Tito and Pappadopulo, Duccio and Ruderman, Joshua T.",
    title = "{Fourth Exception in the Calculation of Relic Abundances}",
    eprint = "1705.08450",
    archivePrefix = "arXiv",
    primaryClass = "hep-ph",
    doi = "10.1103/PhysRevLett.119.061102",
    journal = "Phys. Rev. Lett.",
    volume = "119",
    number = "6",
    pages = "061102",
    year = "2017"
}

@article{Junius:2019dci,
    author = "Junius, Sam and Lopez-Honorez, Laura and Mariotti, Alberto",
    title = "{A feeble window on leptophilic dark matter}",
    eprint = "1904.07513",
    archivePrefix = "arXiv",
    primaryClass = "hep-ph",
    reportNumber = "ULB-TH/19-01",
    doi = "10.1007/JHEP07(2019)136",
    journal = "JHEP",
    volume = "07",
    pages = "136",
    year = "2019"
}

@article{Griest:1990kh,
    author = "Griest, Kim and Seckel, David",
    title = "{Three exceptions in the calculation of relic abundances}",
    reportNumber = "CFPA-TH-90-001A, BA-90-79",
    doi = "10.1103/PhysRevD.43.3191",
    journal = "Phys. Rev. D",
    volume = "43",
    pages = "3191--3203",
    year = "1991"
}

@article{Raj:2024guv,
    author = "Raj, Nirmal and Mondal, Biprajit",
    title = "{Neutrino roof: Single-scatter cross section ceilings in dark matter direct detection experiments}",
    eprint = "2406.17015",
    archivePrefix = "arXiv",
    primaryClass = "hep-ph",
    doi = "10.1103/PhysRevD.110.095023",
    journal = "Phys. Rev. D",
    volume = "110",
    number = "9",
    pages = "095023",
    year = "2024"
}

@article{Bhattacharya:2017sml,
    author = "Bhattacharya, Subhaditya and Sahoo, Nirakar and Sahu, Narendra",
    title = "{Singlet-Doublet Fermionic Dark Matter, Neutrino Mass and Collider Signatures}",
    eprint = "1704.03417",
    archivePrefix = "arXiv",
    primaryClass = "hep-ph",
    doi = "10.1103/PhysRevD.96.035010",
    journal = "Phys. Rev. D",
    volume = "96",
    number = "3",
    pages = "035010",
    year = "2017"
}

@article{Lee:2026jxl,
    author = "Lee, Seung J. and Youn, Taewook",
    title = "{Mixing-suppressed inelastic dark matter: a minimal model for the LZ 248 keV event}",
    eprint = "2609.09138",
    archivePrefix = "arXiv",
    primaryClass = "hep-ph",
    month = "9",
    year = "2026"
}

@article{Zhu:2026dag,
    author = "Zhu, Pengxuan and Dalla Valle Garcia, Giovani and Wang, Xuan-Gong and Thomas, Anthony W. and White, Martin J.",
    title = "{Endothermic dark matter with a light dark photon and the LUX-ZEPLIN high-energy nuclear-recoil candidate}",
    eprint = "2609.09015",
    archivePrefix = "arXiv",
    primaryClass = "hep-ph",
    reportNumber = "ADP-26-15/T1312",
    month = "9",
    year = "2026"
}

@article{Jaegle:2015fme,
    author = "Jaegle, I.",
    collaboration = "Belle",
    title = "{Search for the dark photon and the dark Higgs boson at Belle}",
    eprint = "1502.00084",
    archivePrefix = "arXiv",
    primaryClass = "hep-ex",
    reportNumber = "Belle-Preprint-2014-20, KEK-Preprint-2014-36",
    doi = "10.1103/PhysRevLett.114.211801",
    journal = "Phys. Rev. Lett.",
    volume = "114",
    number = "21",
    pages = "211801",
    year = "2015"
}

@article{BaBar:2012bkw,
    author = "Lees, J. P. and others",
    collaboration = "BaBar",
    title = "{Search for Low-Mass Dark-Sector Higgs Bosons}",
    eprint = "1202.1313",
    archivePrefix = "arXiv",
    primaryClass = "hep-ex",
    reportNumber = "BABAR-PUB-11-025, SLAC-PUB-14863",
    doi = "10.1103/PhysRevLett.108.211801",
    journal = "Phys. Rev. Lett.",
    volume = "108",
    pages = "211801",
    year = "2012"
}
\end{document}